\documentclass[letterpaper,11pt]{article}
\pdfoutput=1

\usepackage{jheppub}
\usepackage{graphicx}
\usepackage{xcolor}
\usepackage[caption=false]{subfig}
\usepackage{amsmath}
\usepackage{amssymb}
\usepackage{comment}
\usepackage{multirow}
\usepackage{longtable}

\usepackage{float}
\hypersetup{
 colorlinks=true,
 citecolor=blue,
 linkcolor=blue,
 urlcolor=blue}

\title{\boldmath Identifying the Quantum Properties \\ of Hadronic Resonances using Machine Learning}
\author[1]{Jakub Filipek,}
\author[1]{Shih-Chieh Hsu,}
\author[1]{John Kruper,}
\author[2]{Kirtimaan Mohan,}
\author[3,4]{and Benjamin Nachman}
\affiliation[1]{\normalsize\it Department of Physics, University of Washington, Seattle, WA 98195, USA}
\affiliation[2]{\normalsize\it Department of Physics and Astronomy Division, Michigan State University, 567 Wilson Rd, East Lansing, MI 48824, USA}
\affiliation[3]{\normalsize\it Physics Division, Lawrence Berkeley National Laboratory, Berkeley, CA 94720, USA}
\affiliation[4]{\normalsize\it Berkeley Institute for Data Science, University of California, Berkeley, CA 94720, USA}

\emailAdd{balbok@uw.edu}
\emailAdd{schsu@uw.edu}
\emailAdd{jk23@att.net}
\emailAdd{kamohan@msu.edu}
\emailAdd{bpnachman@lbl.gov}

\preprint{MSUHEP-21-009}

\begin{document}

\abstract{
	With the great promise of deep learning, discoveries of new particles at the Large Hadron Collider (LHC) may be imminent.
	Following the discovery of a new Beyond the Standard model particle in an all-hadronic channel, deep learning can also be used to identify its quantum numbers. Convolutional neural networks (CNNs) using jet-images can significantly improve upon existing techniques to identify the quantum chromodynamic (QCD) (`color') as well as the spin of a two-prong resonance using its substructure.
	Additionally, jet-images are useful in determining what information in the jet radiation pattern is useful for classification, which could inspire future taggers.
	These techniques improve the categorization of new particles and are an important addition to the growing jet substructure toolkit, for searches and measurements at the LHC now and in the future.
}

\maketitle

\clearpage

\section{\label{sec:intro}Introduction}

Many theories of physics beyond the Standard Model (BSM) predict new particles that exclusively or at least predominantly decay hadronically.   As a result, the experiments at the Large Hadron Collider (LHC) have dedicated significant resources to search for such particles across a wide kinematic range.  If the masses of the new particles are at or below the electroweak scale, then they may be produced with significant Lorentz boost.  Both ATLAS~\cite{Aaboud:2018zba,Aaboud:2018eoy,ATLAS-CONF-2018-052,Aaboud:2017ecz,Aad:2020cws} and CMS~\cite{Sirunyan:2017dnz,Sirunyan:2017nvi,Sirunyan:2019sgo,Sirunyan:2018ikr} have performed searches for direct or indirect (via a cascade decay) boosted BSM resonances that are fully contained inside a single large-radius jet.  If such a particle was to be discovered, a critical task will be to identify its quantum numbers, such as the spin, electric charge and color state. 

The goal of this paper is to investigate the potential of tagging the quantum numbers of hadronically decaying resonances using all of the available information inside the resulting jet.  There are now many applications of deep learning to jet substructure~\cite{Larkoski:2017jix} with a variety of jet representations used for the learning (images~\cite{deOliveira:2015xxd,Almeida:2015jua,Komiske2017b,Baldi:2016fql,Barnard2017b,Kasieczka:2017nvn,Lin:2018cin,Chung:2020ysf}, ordered lists~\cite{Guest:2016iqz,ATL-PHYS-PUB-2017-003,CMS-DP-2018-058,Pearkes:2017hku,Egan:2017ojy}, trees~\cite{Louppe2017b,Cheng:2017rdo}, graphs~\cite{henrion,Qu:2019gqs,Moreno:2019bmu,Bernreuther:2020vhm,Mikuni:2020wpr,Dreyer:2020brq}, unordered sets~\cite{Komiske:2018cqr,Mikuni:2021pou,Dolan:2020qkr}, observable bases~\cite{Datta:2017rhs,Datta:2017lxt,Datta:2019ndh,Komiske:2017aww,Chakraborty:2020yfc}, etc.~\cite{Butter:2017cot,Chakraborty:2019imr,Dreyer:2018nbf} - see Ref.~\cite{2102.02770} for additional references). Jet images~\cite{Cogan:2014oua} are used here for their interpretability.

Non-deep learning observables have already been used by the LHC experiments to identify quantum properties of hadronic resonances~\cite{Asquith:2018igt,Larkoski:2017jix,Altheimer:2012mn,Altheimer:2013yza,Marzani:2019hun,Abdesselam:2010pt}.  For example, variations of the jet charge~\cite{Field:1977fa,Krohn:2012fg} have been used to study quark and gluon jets~\cite{Aad:2015cua,Sirunyan:2017tyr} as well as boosted hadronically decaying $W$ and $Z$ bosons~\cite{Aad:2015eax}.  These techniques were recently combined with deep learning approaches~\cite{Fraser:2018ieu,Chen:2019uar} to show that additional discrimination power is possible when all of the charge-sensitive information is utilized.  Another quantum property which has been well studied with non-deep learning techniques is the color representation.  For example, the jet pull~\cite{Gallicchio:2010sw} has been used to study singlet versus octet radiation patterns from dijet resonances~\cite{Abazov:2011vh,Aad:2015lxa,Aaboud:2018ibj}.  So far, there has not been a corresponding deep learning study to identify potential gains from using all of the available information in the jet radiation pattern.

The authors of Ref.~\cite{Chivukula:2017nvl} performed a first study to categorize various boosted hadronic resonances based on their quantum numbers.  By using classical jet substructure observables, they were able to show well one can distinguish various possibilities\footnote{Recently, Ref.~\cite{Buckley:2020kdp} also showed that in the color singlet case, one can derive new observables using analytic insight.}.  The goal of this paper is to extend the analysis of Ref.~\cite{Chivukula:2017nvl} using deep learning techniques and quantify the pairwise distinguishability of BSM scenarios.   Deep learning approaches are compared with more traditional methods and various visualizations are used to identify the useful information within the jet radiation pattern.  While the jet-by-jet classification performance is weak, even a small amount of separation can be used in a template fit to extract the quantum numbers.  Thus, we envision that this approach will be used in a post-discovery phase to categorize the resonance without needing excellent per-event distinguishing power.

This paper is organized as follows.  Section~\ref{sec:sim} introduces the BSM models and how they are simulated.  The machine learning setup is documented in Sec.~\ref{sec:ml}, starting with the jet image preparation and preprocessing.  Results are presented in Sec.~\ref{sec:results} and the dependence of these results on variations in the setup is explored in Sec.~\ref{sec:dep}.  The paper ends with conclusions and future outlook in Sec.~\ref{sec:concl}.

\section{\label{sec:sim}Simulation Setup}

A set of leptohobic resonances are used as benchmarks and are described below.  In all cases, the mass of the new resonance is set to 125 GeV.  The model UFO files are generated using {\tt FeynRules}~\cite{Alloul:2013bka}. In the list below we describe the different resonances and their interaction with standard model particles. We denote quark and and anti-quark fields by $q$ and $\bar q$ respectively, the strong coupling constant by $g_s$ and  the gluonic field strength tensor as $G^{a}_{\mu\nu}$.

\begin{enumerate}
	\item New gauge bosons such as Z-primes ($Z^{\prime}$) are a common occurrence in models with extended electro-weak gauge sectors. The benchmark considered here is a color singlet leptophobic $Z^{\prime}$. The $Z^{\prime}$ interacts only with quarks through $\frac{g_B}{6}\bar{q}\gamma^{\mu}q Z^{\prime}_{\mu}$~\cite{Dobrescu:2013coa}. Here $g_B$ is the $Z^{\prime}$ gauge coupling and determines the strength of its interactions to quarks.

	\item New massive colored gauge bosons such as colorons ($C_{\mu}$) are common in models with an extended gauge symmetry of the strongly interacting sector. The color octet colorons interact with quarks through $g_s \tan{\theta}\ \bar{q}\  T^{a}\gamma^{\mu}q C^{a}_{\mu}$\cite{Dobrescu:2013coa,Chivukula:1996yr}. Here $\theta$ is a dimensionless parameter that parametrizes mixing between the gauge and mass eigenbasis of the gluon and coloron.

	\item Scalar diquarks appear in the 27-dimensional representation of string inspired grand-unified theories~\cite{Hewett:1988xc}. The benchmark in this paper is a color sextet diquark $(\Phi^\gamma_6)$ that decays to  pairs of quarks through $\sqrt{2} (\bar{K}_6)^{ab}_\gamma \lambda_\Phi \Phi^\gamma_6 \bar{u}_{Ra}u_{Lb} $\cite{Chivukula:2015zma}. Here $(\bar{K}_6)^{ab}$ is the relevant $SU(3)_c$ Clebsch-Gordon coefficient and $\lambda_\Phi$ is a coupling constant.

	\item In composite models, quarks can be composite at a scale $\Lambda$. As a result  color triplet excited quarks ($q^*$) are a prediction of such models and interact with quarks and gluons (as well as other gauge bosons) through the interaction term
	$\frac{1}{2\Lambda} \bar{q}^{*}_R \sigma^{\mu\nu} [ g_S f_S \frac{\lambda^{a}}{2} G^{a}_{\mu\nu}]q_{L}$  \cite{Baur:1987ga}. Here $q_L$ represents the doublet of left handed Standard Model quarks, $\lambda^a$ are the Gell-Mann matrices, $\sigma^{\mu\nu}$ is the usual bilinear covariant and $f_s$ is a coupling contant and $\Lambda$ is the scale of validity of the effective field theory.

	\item Massive Spin 2 objects ($X^{\mu\nu}$) may appear as Kaluza-Klein excitations of the gravitational field $h^{\mu\nu}$ in RS models. They are color singlets and interact with SM particles through the energy momentum tensor $T_{\mu\nu}$ as $\frac{1}{\Lambda}X^{\mu\nu}T_{\mu\nu}$ \cite{Han:1998sg,Chivukula:2020hvi}, decaying to either quarks or gluons. Here $\Lambda$ represents the scale of validity of the effective field theory~\cite{Chivukula:2020hvi,Chivukula:2019rij,SekharChivukula:2021btd,Chivukula:2019zkt}.
	
	\item Color octet scalars ($S_{8}$) that are electroweak singlets may arise in technicolor or extra-dimension models~\cite{Hill:2002ap,Dobrescu:2007yp}. They interact with gluons through the field strength tensor as  $\frac{g_s d_{ABC} k_s}{\Lambda} S_{8}^A G_{\mu\nu}^{B}G^{C,\mu\nu}$. Here $d_{ABC}$ are the usual $SU(3)$ structure constants and $\Lambda$ represents the scale of validity of the effective field theory.
	\item Two additional models are color singlet scalars that decay to a pair of gluons or a pair of quarks, like the SM Higgs boson.  These particles are denoted by $H\rightarrow qq$ or $H\rightarrow gg$.
\end{enumerate}

Proton-proton collisions involving the above models are generated with {\tt MadGraph5\_amc@NLO}~\cite{Alwall:2011uj},
version 2.5.5, and {\tt Pythia8}~\cite{Sjostrand2008a}, version 8.235.  Particles with a mean lifetime $\tau>30$ ps, excluding neutrinos, are clustered into jets with {\tt FastJet}~3~\cite{Cacciari2012a,Cacciari:2005hq}, version 3.3.2, using the anti-$k_t$ algorithm~\cite{Cacciari2008b} with R = 1.0 and are required to be central with a pseudorapidity $\vert\eta\vert < 2.0$.  Due to the benchmark mass of the resonance, the leading jet $p_T$ is required to be between $300$~GeV and $600$~GeV to ensure that one large-radius jet is sufficient to capture most of the hadronic decay products.  As in Ref.~\cite{deOliveira:2015xxd}, jets are trimmed~\cite{Krohn:2009th} by re-clustering the constituents into $R = 0.3$ $k_t$ subjets and and dropping those which have $p_T^\text{subjet} < 0.05 \times p_T^\text{jet}$.  Jet grooming is revisited in Sec.~\ref{sec:dep}.  In order for an event to be used for machine learning, the leading jet must have $m_\text{jet}\in[100,150]$ GeV.

As stated in Sec.~\ref{sec:intro}, the premise of this study is that a new resonance is identified and an event sample is investigated to identify its properties.   The task of identifying the BSM in the first place from the generic quark and gluon jet background is beyond the scope of this goal.

\section{\label{sec:ml}Machine Learning}

\subsection{\label{sec:preML}Preprocessing}

Jet images are constructed and preprocessed using a similar methodology as the one from Ref.~\cite{deOliveira:2015xxd} and briefly reviewed in the following.  Images cover a range in $\eta$-$\phi$ that is $2R \times 2R$ where $R = 1$ is the radius of the jet.  Jets are translated so that the average weighted on pixel intensity is at the origin.  The images are pixelated with a size $65 \times 65$.  The second subjet is used to rotate the jet so that the vector connecting the leading and subleading subjets is at $-\pi/2$.  In the absence of a second subjet, the principle component axes are used instead.  A last degree of freedom is removed by flipping the images horizontally so that the right side of the image has a higher sum $p_T$ than the left side.  The pixel intensities are then transformed with a logarithm and normalized to unity.  The lossy nature of these preprocessing steps has been studied elsewhere~\cite{deOliveira:2015xxd,deOliveira:2017pjk}.

\begin{figure}[H]
	\centering
	\includegraphics[scale=0.9]{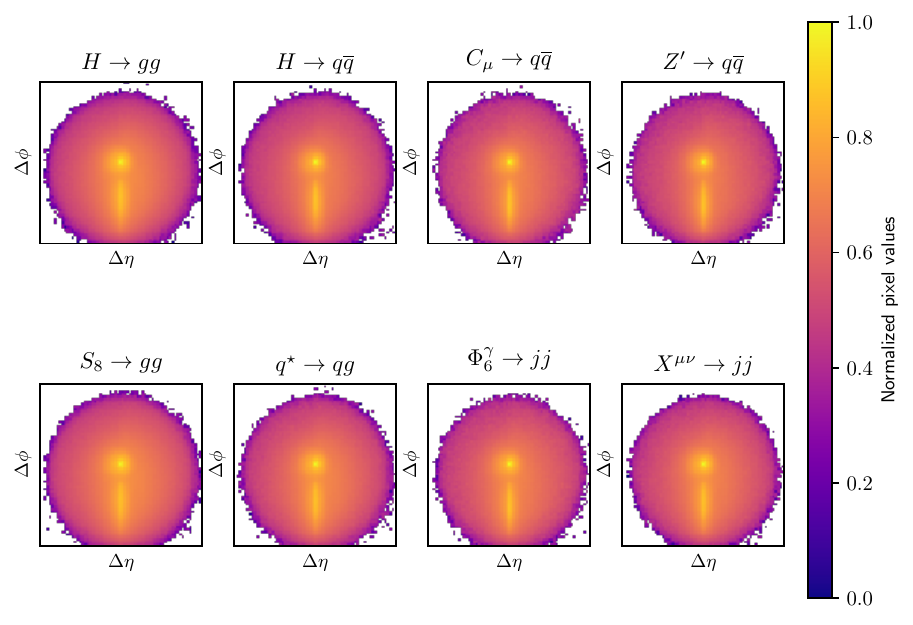}
	\includegraphics[scale=0.9]{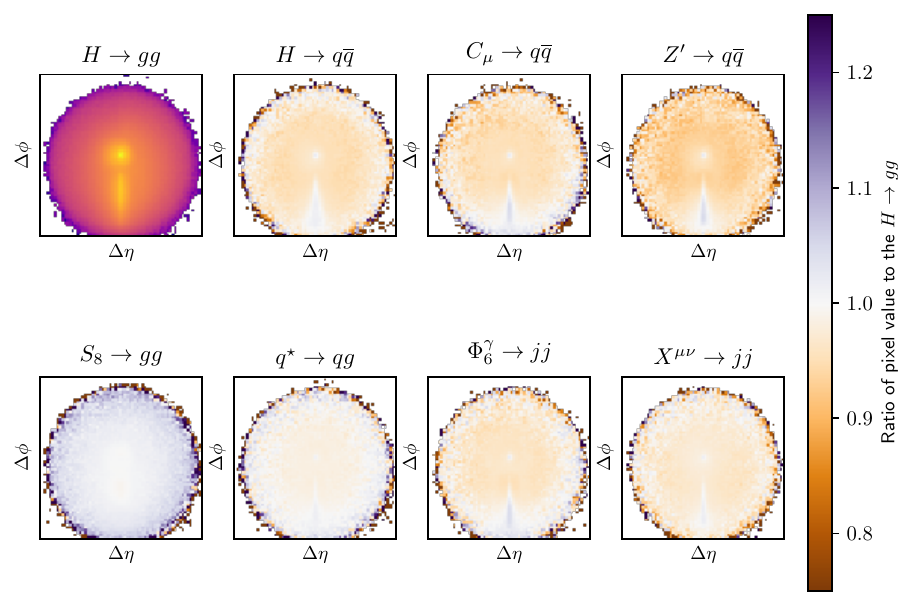}
	\caption{Top: Average images of jets generated by the processes considered in Sec.~\ref{sec:sim}. Bottom: Per pixel ratios of the average images to the $H \rightarrow gg$ case. The upper-left image is left for reference, since the ratio of $H \rightarrow gg$ with itself would be all ones. Decays with the same final state have more similar jet images. While most processes go to either $q\bar{q}$ or $gg$, the excited quark ($q^\star$) goes to $qg$ (or $\bar{q}g$). In this paper, classifiers are trained to distinguish these processes from each other.}
    \label{fig:jetimage}
\end{figure}

The average jet image for the event samples described in Sec.~\ref{sec:sim} is presented in Fig.~\ref{fig:jetimage}.  The leading subjet is the bright spot in the center of the image and the elongated bright spot below it corresponds to the subleading subjet.  All eight of the average jet images in the top plot of Fig.~\ref{fig:jetimage} look similar by eye and so the bottom plot shows the ratio of the average jet image for a given sample to the one from the $H\rightarrow g g$ sample.  The ratio images show systematic trends at the 5-10\% level that illustrate the information that a neural network could use for distinguishing the various possibilities.

\subsection{\label{sec:CNNml}Low Level Tagger: Convolutional Neural Network on Jet Images}
\label{sec:CNNDescription}

\begin{figure}[H]
	\centering
	\includegraphics[scale=0.5]{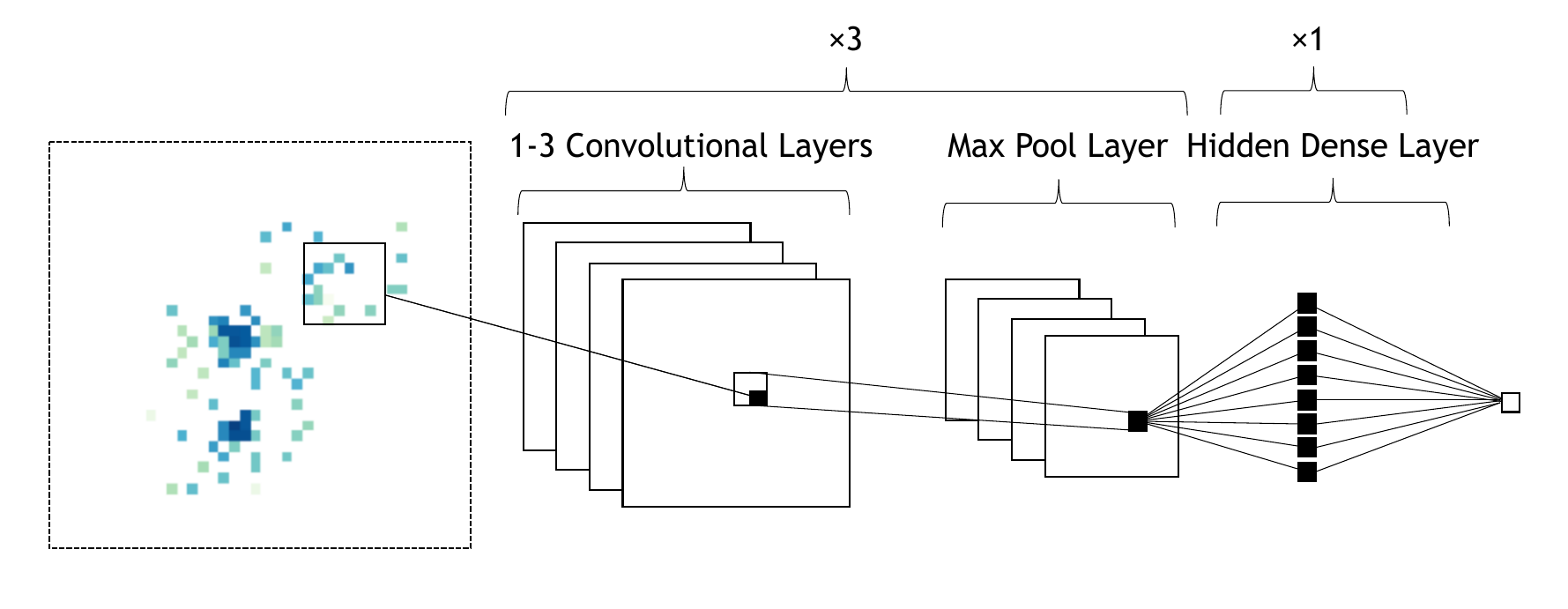}
	\caption{Architecture of the convolutional neural network. The input is a preprocessed $65\times65$ jet-image. There are eight convolutional layers with three intermixed max-pooling layers, followed by a dense layer. The output is a single number.}
	\label{fig:architecture}
\end{figure}

A Convolutional Neural Network (CNN) (Fig.~\ref{fig:architecture}) consisting of  eight layers of convolutional filters, one dense layer with 128 neurons, and a final dense layer with a single output neuron is used. The first convolutional layer has eight $11\times11$ filters, all other convolutional layers have sixteen $3\times3$ filters. Each of the third, fifth, and seventh convolutional layers is followed by $2\times2$ max-pooling layers. The hidden dense layer is followed by a dropout layer with 50\% dropout. All layers are followed by rectified linear unit (ReLU) activation. \newline \null

The CNN training minimized the binary cross-entropy loss using the Nesterov Adam optimizer~\cite{Kingma2014, Sutskever:2013:IIM:3042817.3043064} with a learning rate of $3\times10^{-4}$ and a scheduled decay of 0.153.  As the radiation pattern of the different models varies on the periphery of the jet, the jet mass distribution is not identical for all models.  To remove the classification power of the jet mass, event are weighted (using fine-binned histograms) for each binary classification task. The training is done with a batch size of 1024 for 200 epochs. The best model according to the validation loss is saved. Keras~\cite{chollet2015keras} built on TensorFlow~\cite{tensorflow2015-whitepaper} is used to train the CNN. \newline

The Sequential Model-based Algorithm Configuration in python 3 (SMAC3) is used to optimize the CNN hyperparameters~\cite{SMAC3}. SMAC3 is based on Bayesian Optimization, and is used to optimize arbitrary algorithms. The hyperparameters considered are the first layer filter size, number of filters per convolutional layer, number of convolutional layers, dropout, activation function, width of the dense layer(s), number of dense layer(s), optimizer type, learning rate, and decay rate. A first layer filter size of $11 \times 11$ is considered because it can capture aspects of both the leading and subleading subjet~\cite{deOliveira:2015xxd}. This is important for learning relations between the sparse, non-zero parts of image. $11$ pixels correspond roughly to the distance between the two most energetic pixels in the $65 \times 65$ image. After the first filter, a series of $3 \times 3$ filters follows, since they are able to handle similar transformations as larger filters without computational overhead~\cite{simonyan2014deep}. The Adam optimizer~\cite{Kingma2014}, the Nesterov Adam optimizer~\cite{Kingma2014, Sutskever:2013:IIM:3042817.3043064}, stochastic gradient descent, and root mean squared propagation (RMSprop) are the CNN optimizers considered. SMAC3 chose the largest number of convolutional layers and the largest dense layer width it considered.

\subsection{\label{sec:EFlow}High Level Tagger: Energy Flow Variables}
Two high level taggers are trained to reduce the feature space and to compare to the CNN performance. For the first tagger, ten high level observables (called `E-flow') are constructed using a method inspired by~\cite{Chien:2017xrb}. 10 linearly spaced rings are defined on the jet image with diameters from 1 to 65 pixels. The pixel activations in each ring are summed. In all of the figures, the observable corresponding to $i^{th}$ ring will be refereed to as $E_i$. AdaBoost~\cite{AdaBoost} from scikit-learn~\cite{scikit-learn} is used as the classifier. Adaboost takes a weighted sum of the outputs of base classifiers to output a final classification. The base classifiers used are decision trees with a maximum depth of 3. A maximum number of 200 decision trees are used.

\begin{figure}[H]
	\centering
	\includegraphics[scale=0.5]{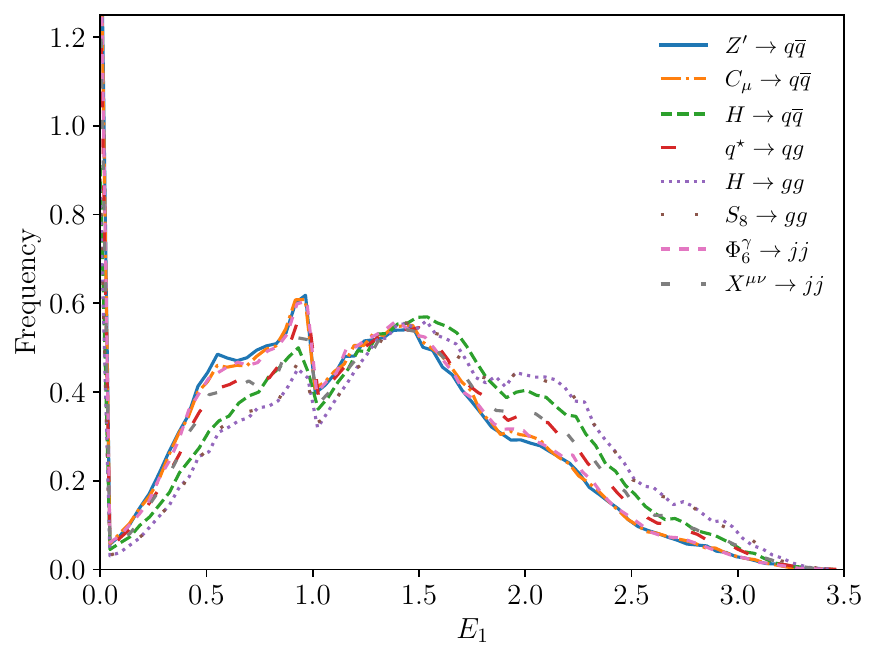}
	\includegraphics[scale=0.5]{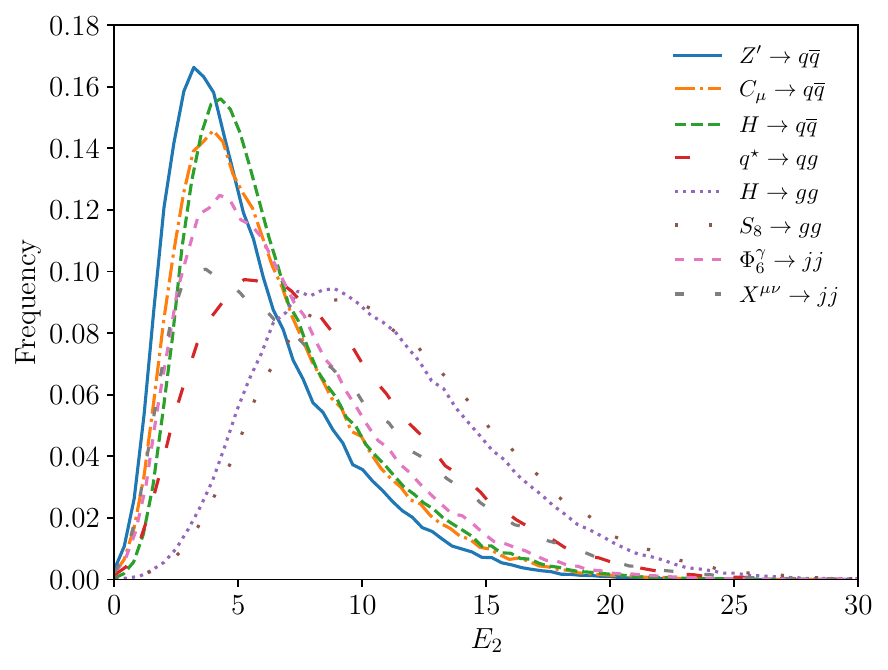}
	\includegraphics[scale=0.5]{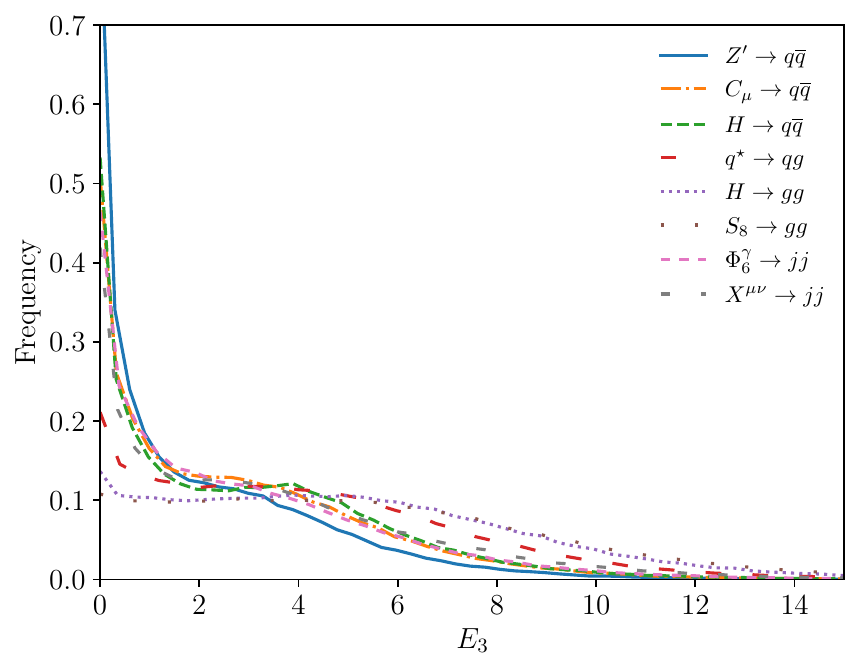}
	\includegraphics[scale=0.5]{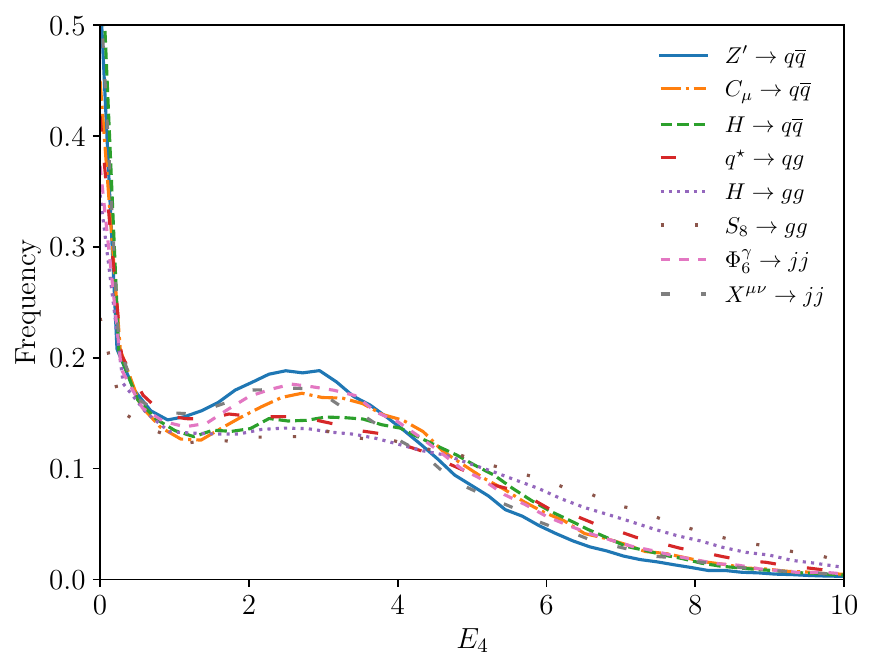}
	\includegraphics[scale=0.5]{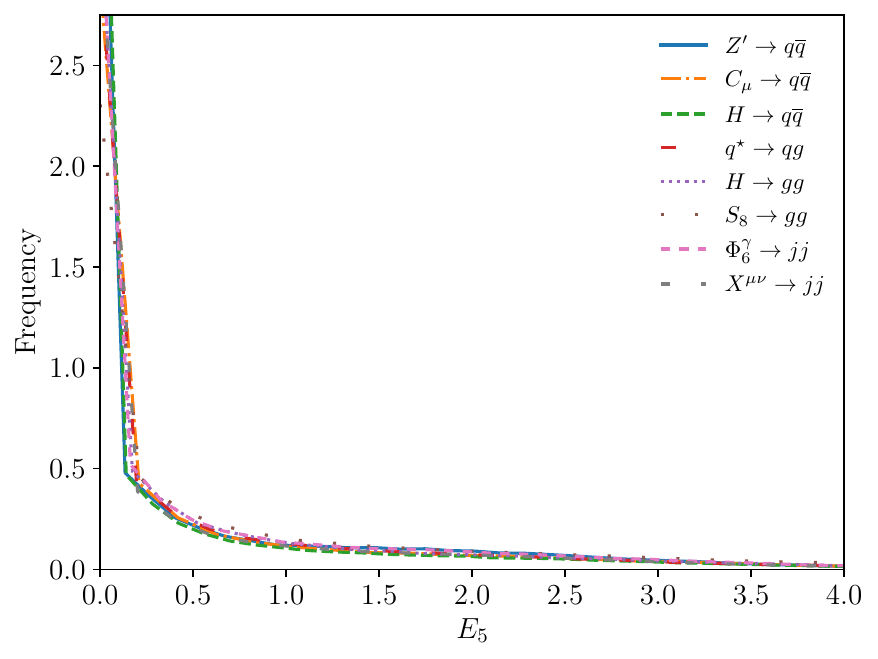}
	\includegraphics[scale=0.5]{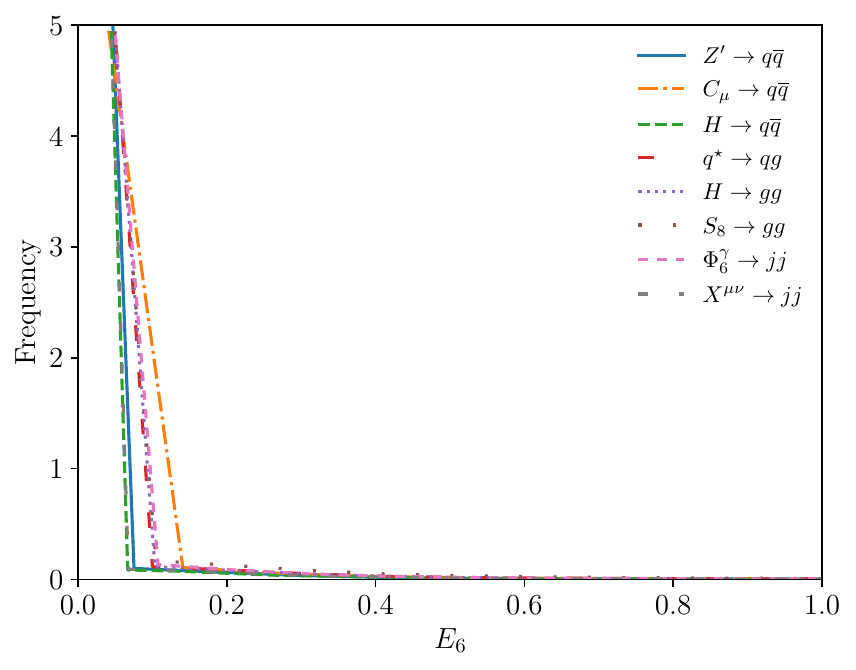}
	\caption{Distributions of the first six E-flow variables normalized to unity. The higher indices correspond to larger radii within the jet. Majority of distributions peak around 0, which shows that most of jets do not contain a lot of sub-jets. It interestingly occurs for $E_1$. Possible reason is that the leading subjet might occur in the pixels surrounding the center one, if the rest of subjets have high enough $p_T$.
 }
	\label{fig:tjethists}
\end{figure}

Example distributions of the $E_i$ observables are presented in Fig.~\ref{fig:tjethists}. While it is possible to see differences between the different types of color resonances in almost all the plots. The most noticeable difference occurs in $E_2$. Here we see how the distribution of $E_2$ peaks closer to zero for resonances that decay to quarks in the final state, such as $Z'$ or $C_\mu$, whereas resonances that decay to gluonic final states, such as $S_8$ peak at larger values of $E_2$. Resonances that decay to a mixture of quarks and gluons, such as $q^\star$ peak at values larger than the purely quark and purely gluonic distributions. As expected, the energy distribution around a gluonic jet is more diffuse, and thus we see that $E_2$ is a good discriminator for distinguishing between quark and gluon final states. Similar patterns can also be observed for $E_3$ and $E_4$, possibly coming from the second subjet. 

\clearpage

\subsection{High Level Tagger: Jet Pull}

As discussed in Sec.~\ref{sec:intro}, the jet pull angle~\cite{Gallicchio:2010sw} is an observable that has been demonstrated to be sensitive to the color representation of the new resonance~\cite{Abazov:2011vh,Aad:2015lxa,Aaboud:2018ibj}.  The jet pull angle is computed from the subjet pull vector, which is a $p_T$ weighted radial moment:

\begin{align}
\sum_{i \in J} \frac{p^i_T|\Vec{r_i}|}{p^J_T}\Vec{r_i}\,,
\end{align}

\noindent where $J$ is the given subjet, $p^J_T$ is its transverse momentum, $i$ iterates over the constituents of $J$, and $\Vec{r_i}$ is the vector difference between the subjet axis and the $i^{th}$ constituent in rapidity-azimuthal angle space ($\Delta r_i = (\Delta y_i, \Delta \phi_i)$).  The pull angle is defined for two subjets $J_1$ and $J_2$ as the angle formed between the pull vector of one of the jets and the vector difference between $J_1$ and $J_2$. Each jet has a corresponding pull angle, $\theta_{J_1,J_2}$ and $\theta_{J_2,J_1}$. \newline \null \quad

\begin{figure}[H]
	\centering
	\includegraphics[scale=0.5]{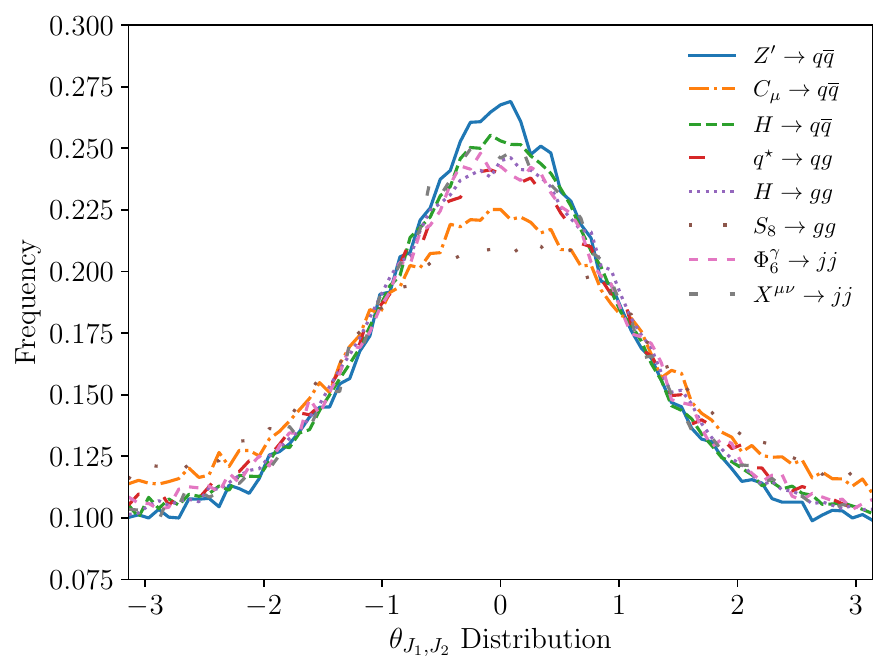}
	\includegraphics[scale=0.5]{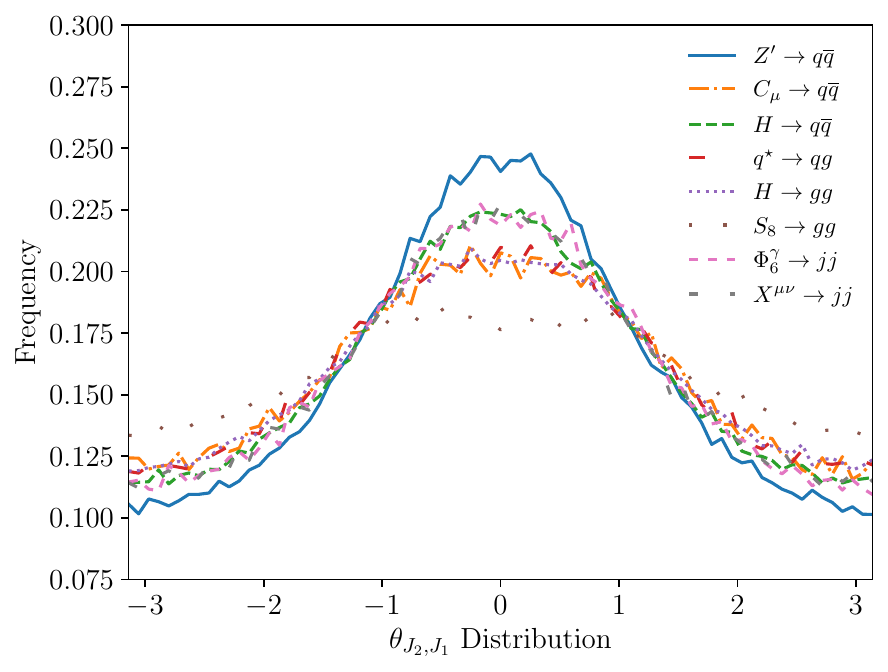}
	\caption{Distributions of both jet pull angles for each process.}
	\label{fig:pullhists}
\end{figure}

The pull angle distribution is predicted to be more uniform when the two quarks 
from the resonance decay are not color-connected to each other and more peaked at zero if they are connected.  The distributions of the two jet pull angles for each of the resonance models are shown in Fig.~\ref{fig:pullhists}.  As expected, the color singlets $Z'\rightarrow q \bar{q}$ and $H\rightarrow q \bar{q}$ have pull angles more closely distributed around 0, while the color octets $C_\mu \rightarrow gg$ and $S_8 \rightarrow gg$ have a more spread out distribution.

For each event, the two pull angles are used as high level features. The two angles are combined using the same  AdaBoost algorithm as described in Sec.~\ref{sec:EFlow}.

\clearpage

\section{\label{sec:results}Results}

For each binary classification task, 300,000 events are used with a train-test split of 80-20\%. For the CNN, the training set is further sub-divided with a training-validation split of 80-20\% and the CNN's loss function is weighted as described in Sec.~\ref{sec:CNNml}.

\subsection{Training Results}
The performance of the various methods and signal models are compared using significance improvement characteristic (SIC) curves and receiver operating characteristic (ROC) curves. For a given signal efficiency (true positive rate) $\epsilon$, SIC curves show ${\epsilon}/{\sqrt{\epsilon_b}}$ and ROC curves show $\epsilon_b$, where $\epsilon_b$ is the background efficiency (false positive rate).  Note that all samples are BSM signals, but in each binary task, one is labeled as `signal' and other as 'background'.  The two classes are treated symmetrically in the loss function so this labeling scheme is irrelevant. 

\begin{figure}[H]
	\centering
	\includegraphics[scale=0.49]{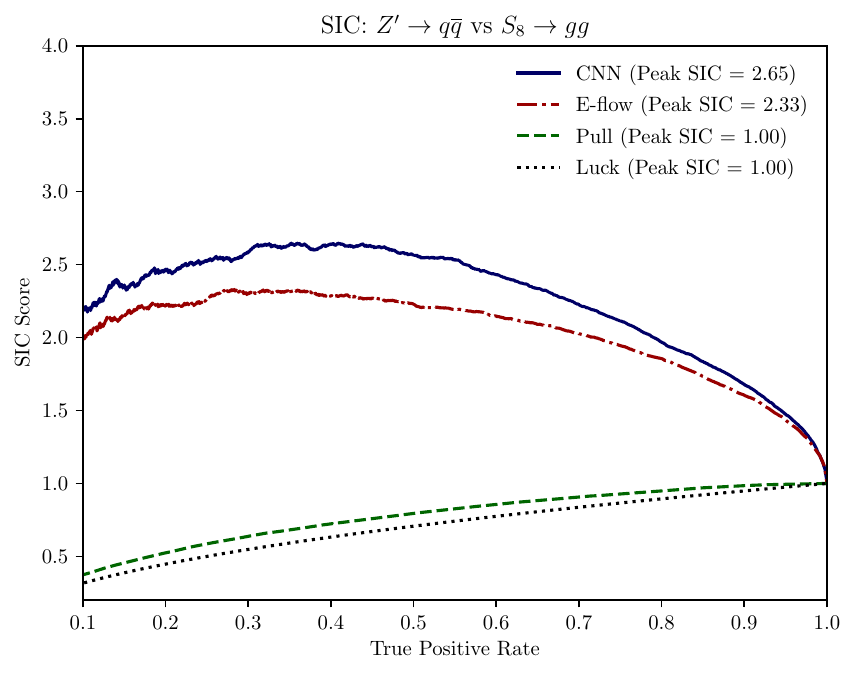}
	\includegraphics[scale=0.49]{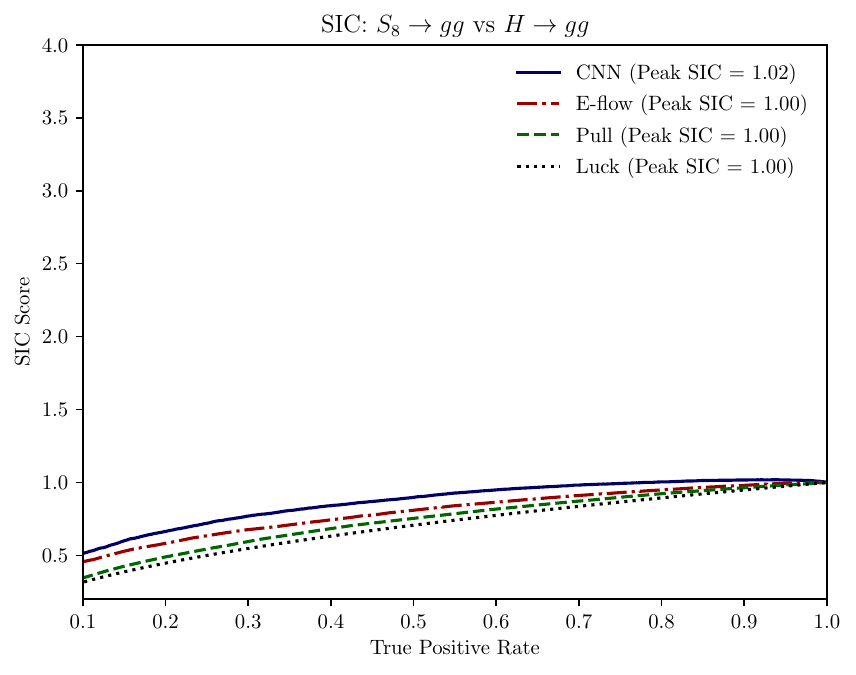}
	\includegraphics[scale=0.49]{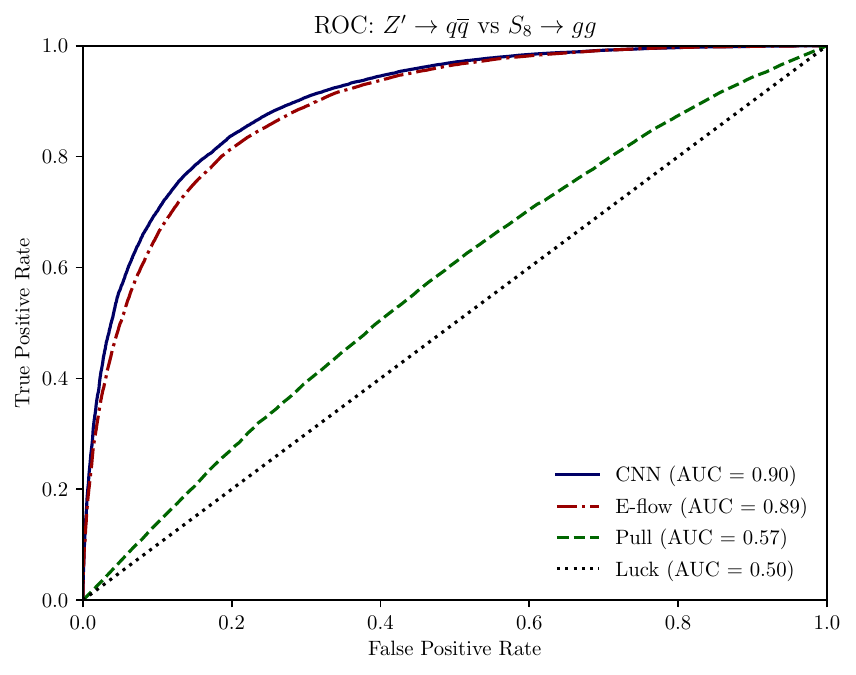}
	\includegraphics[scale=0.49]{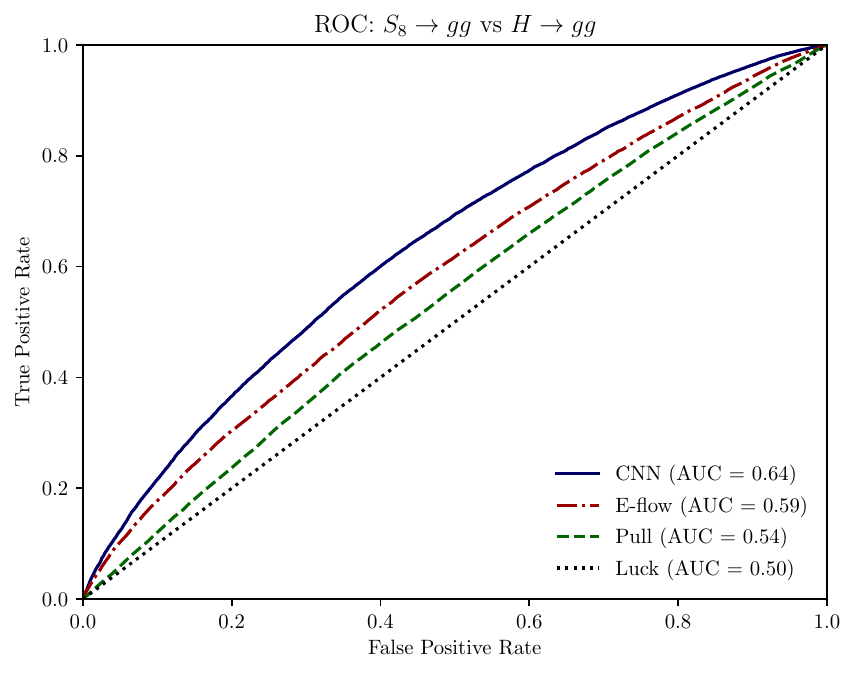}
	\caption{Significance improvement characteristic (SIC) curves (top) and Receiver Operating Characteristic (ROC) curves (bottom).  CNN represents the convolutional neural network trained on the low level jet images. E-flows represents AdaBoost trained on the high level telescoping jet features. Pull represents AdaBoost trained on the high level jet pull angle angle features. Luck represents the scores that would be achieved by a random classifier.  The classification task is $Z' \rightarrow q\bar{q}$ from $S_8 \rightarrow gg$ on the left and $S_8 \rightarrow gg$ from $H \rightarrow gg$ on the right.}
	\label{fig:siccurves}
\end{figure}

Figure~\ref{fig:siccurves} shows SIC and ROC curves for the three taggers in two different signal comparisons. The CNN has the strongest performance, followed by the E-flows and the jet pull. The curves on the left show stronger discrimination power than the ones on the right. This is expected, since the final state for the two BSM particles are  different for the plots on the left and the same for the curves on the right. 
We also quote the peak SIC value, which is the highest value attained in the SIC curve, and  area under curve (AUC), which  is the area under the ROC curve. Both the peak SIC and the AUC can be used as single values which quantify performance.

We summarize the performance of the CNN and compare it to the performance of other high level taggers using the peak SIC and AUC values in  Table~\ref{tab:siccompare}. We see that the CNN performs just as well or better than the other observables in every combination of signals. The samples are ordered such that samples with similar final states are close together. The E-flow observables and CNN perform well when the final states are different. Jet pull is not a strong classifier for any combination. 
In general, it is easier to distinguish BSM particles which decay to pairs of gluons more easily from those that decay into quarks. On the other hand, BSM particles that decay into similar final states are harder to distinguish. For example,  the AUC values of the $C_\mu \rightarrow q \bar{q}$ versus  $H \rightarrow q \bar{q}$ are  lower compared to $C_\mu \rightarrow q \bar{q}$ versus $H \rightarrow g g$. The AUC values further reduce when we compare BSM particles that decay into a similar final state  (quarks or gluons) and also have the same spin quantum numbers. For example,  $C_\mu \rightarrow q \bar{q}$ versus  $Z' \rightarrow q \bar{q}$  has lower AUC values than $C_\mu \rightarrow q \bar{q}$ versus  $H \rightarrow q \bar{q}$. $C_\mu$ and $Z'$ are both vector resonances, but differ in the color quantum numbers, the former being a color octet and the later a color singlet. The only distinguishing feature therefore the small differences in radiation patterns that arise due to different color flow in these two samples. Similar observations can be made about $H \to gg $ versus $S_8 \to gg$ which differ only in their color representation. Finally we note that for the spin-2 resonance $X^{\mu\nu}$, the final state comprises of $X^{\mu\nu} \to gg$ and $X^{\mu\nu} \to q \bar q$ in roughly equal proportions. As expected this degrades the performance of the CNN, also indicating that the CNN performs better discriminating different final states, but not as well with distinguishing the spin quantum numbers of the BSM resonance. 

\begin{center}
	\begin{table}[!htb]
		\resizebox{\textwidth}{!}{
			\begin{tabular}{|c||c||c|c||c|c||c|c||c|c||c|c||c|c||c|c||c|c|}
				\hline
				& Model & \multicolumn{2}{c||}{$Z\prime \rightarrow q\overline{q}$} & \multicolumn{2}{c||}{$C_\mu \rightarrow q\overline{q}$} & \multicolumn{2}{c||}{$H \rightarrow q\overline{q}$} & \multicolumn{2}{c||}{$q^\star \rightarrow qg$} & \multicolumn{2}{c||}{$H \rightarrow gg$} & \multicolumn{2}{c||}{$S_8 \rightarrow gg$} & \multicolumn{2}{c||}{$\Phi^\gamma_6 \rightarrow jj$} & \multicolumn{2}{c|}{$X^{\mu\nu} \rightarrow jj$} \\ 
				\hline
				\multirow{3}{*}{$Z\prime \rightarrow q\overline{q}$}
				& CNN     & SIC                                  & AUC                                  & \textbf{1.05}  & \textbf{0.67}  & \textbf{1.16}  & \textbf{0.72}  & \textbf{1.36}  & \textbf{0.78}  & \textbf{2.31}  & \textbf{0.89}  & \textbf{2.65}  & \textbf{0.90}  & \textbf{1.12} & \textbf{0.71} & \textbf{1.02}  & \textbf{0.67}  \\ 
				& E-flow  &                                      &                                      & 1.00              & 0.61              & 1.89              & 0.86              & 1.27              & 0.76              & 1.89              & 0.86              & 2.33              & 0.89              & 1.02             & 0.64             & 1.02              & 0.67              \\ 
				& Pull    &                                      &                                      & 1.00              & 0.54              & 1.00              & 0.53              & 1.00              & 0.53              & 1.00              & 0.53              & 1.00              & 0.57              & 1.00             & 0.53             & 1.00              & 0.52              \\ 
				\hline 
				\multirow{3}{*}{$C_\mu \rightarrow q\overline{q}$}
				& CNN     & \textbf{1.05}  & \textbf{0.67}  & SIC                                  & AUC                                  & \textbf{1.24}  & \textbf{0.76}  & \textbf{1.24}  & \textbf{0.76}  & \textbf{1.73}  & \textbf{0.85}  & \textbf{2.03}  & \textbf{0.88}  & \textbf{1.00} & \textbf{0.62} & \textbf{1.03}  & \textbf{0.66}  \\ 
				& E-flow  & 1.00              & 0.61              &                                      &                                      & 1.01              & 0.62              & 1.07              & 0.68              & 1.49              & 0.81              & 1.70              & 0.84              & 1.00             & 0.57             & 1.01              & 0.63              \\ 
				& Pull    & 1.00              & 0.54              &                                      &                                      & 1.00              & 0.54              & 1.00              & 0.52              & 1.00              & 0.52              & 1.00              & 0.53              & 1.00             & 0.52             & 1.00              & 0.53              \\ 
				\hline
				\multirow{3}{*}{$H \rightarrow q\overline{q}$}
				& CNN     & \textbf{1.16}   & \textbf{0.72}   & \textbf{1.24}   & \textbf{0.76}   & SIC                                 & AUC                                 & \textbf{1.29}   & \textbf{0.77}   & \textbf{1.73}   & \textbf{0.85}   & \textbf{2.20}   & \textbf{0.89}   & \textbf{1.37}  & \textbf{0.78}  & \textbf{1.13}   & \textbf{0.72}   \\ 
				& E-flow  & 1.04               & 0.65               & 1.01               & 0.62               &                                     &                                     & 1.12               & 0.71               & 1.44               & 0.81               & 1.74               & 0.84               & 1.04              & 0.65              & 1.03               & 0.65               \\ 
				& Pull    & 1.00               & 0.52               & 1.00               & 0.54               &                                     &                                     & 1.00               & 0.52               & 1.00               & 0.52               & 1.00               & 0.56               & 1.00              & 0.51              & 1.00               & 0.51               \\ 
				\hline
				\multirow{3}{*}{$q^\star \rightarrow qg$}
				& CNN     & \textbf{1.36}  & \textbf{0.78}  & \textbf{1.24}  & \textbf{0.76}  & \textbf{1.29}  & \textbf{0.77}  & SIC                                  & AUC                                  & \textbf{1.05}  & \textbf{0.68}  & \textbf{1.12}  & \textbf{0.71}  & \textbf{1.04} & \textbf{0.68} & \textbf{1.03}  & \textbf{0.67}  \\ 
				& E-flow  & 1.27              & 0.76              & 1.07              & 0.68              & 1.12              & 0.71              &                                      &                                      & 1.02              & 0.67              & 1.12              & 0.72              & 1.02             & 0.64             & 1.00              & 0.63              \\ 
				& Pull    & 1.00              & 0.53              & 1.00              & 0.52              & 1.00              & 0.52              &                                      &                                      & 1.00              & 0.50              & 1.00              & 0.54              & 1.00             & 0.51             & 1.00              & 0.51              \\ 
				\hline 
				\multirow{3}{*}{$H \rightarrow gg$}
				& CNN     & \textbf{2.31}   & \textbf{0.89}   & \textbf{1.73}   & \textbf{0.85}   & \textbf{1.73}   & \textbf{0.85}   & \textbf{1.05}   & \textbf{0.68}   & SIC                                 & AUC                                 & \textbf{1.02}   & \textbf{0.64}   & \textbf{1.59}  & \textbf{0.79}  & \textbf{2.20}   & \textbf{0.73}   \\ 
				& E-flow  & 1.89               & 0.86               & 1.49               & 0.81               & 1.44               & 0.81               & 1.02               & 0.67               &                                     &                                     & 1.00               & 0.59               & 1.52              & 0.78              & 1.58               & 0.72               \\ 
				& Pull    & 1.00               & 0.53               & 1.00               & 0.52               & 1.00               & 0.52               & 1.00               & 0.50               &                                     &                                     & 1.00               & 0.54               & 1.00              & 0.51              & 1.00               & 0.52               \\ 
				\hline
				\multirow{3}{*}{$S_8 \rightarrow gg$}
				& CNN     & \textbf{2.65}  & \textbf{0.90}  & \textbf{2.03}  & \textbf{0.88}  & \textbf{2.20}  & \textbf{0.89}  & \textbf{1.12}  & \textbf{0.71}  & \textbf{1.02}  & \textbf{0.64}  & SIC                                  & AUC                                  & \textbf{1.80} & \textbf{0.81} & \textbf{2.76}  & \textbf{0.78}  \\ 
				& E-flow  & 2.33              & 0.89              & 1.70              & 0.84              & 1.74              & 0.84              & 1.12              & 0.72              & 1.00              & 0.59              &                                      &                                      & 1.86             & 0.81             & 2.16              & 0.76              \\ 
				& Pull    & 1.00              & 0.57              & 1.00              & 0.53              & 1.00              & 0.56              & 1.00              & 0.54              & 1.00              & 0.54              &                                      &                                      & 1.00             & 0.55             & 1.00              & 0.55              \\ 
				\hline
				\multirow{3}{*}{$\Phi^\gamma_6 \rightarrow jj$}
				& CNN     & \textbf{1.12} & \textbf{0.71} & \textbf{1.00} & \textbf{0.62} & \textbf{1.37} & \textbf{0.78} & \textbf{1.04} & \textbf{0.68} & \textbf{1.59} & \textbf{0.79} & \textbf{1.80} & \textbf{0.81} & SIC                                  & AUC                                  & \textbf{1.06} & \textbf{0.68} \\ 
				& E-flow  & 1.02             & 0.64             & 1.00             & 0.57             & 1.04             & 0.65             & 1.02             & 0.64             & 1.52             & 0.78             & 1.86             & 0.81             &                                      &                                      & 1.01             & 0.61             \\ 
				& Pull    & 1.00             & 0.53             & 1.00             & 0.52             & 1.00             & 0.51             & 1.00             & 0.51             & 1.00             & 0.51             & 1.00             & 0.55             &                                      &                                      & 1.00             & 0.51             \\ 
				\hline
				\multirow{3}{*}{$X^{\mu\nu} \rightarrow jj$}
				& CNN     & \textbf{1.02}  & \textbf{0.67}  & \textbf{1.03}  & \textbf{0.66}  & \textbf{1.13}  & \textbf{0.72}  & \textbf{1.03}  & \textbf{0.67}  & \textbf{2.20}  & \textbf{0.73}  & \textbf{2.76}  & \textbf{0.78}  & \textbf{1.06} & \textbf{0.68} & SIC                                  & AUC                                  \\ 
				& E-flow  & 1.02              & 0.67              & 1.01              & 0.63              & 1.03              & 0.65              & 1.00              & 0.63              & 1.58              & 0.72              & 2.16              & 0.76              & 1.01             & 0.61             &                                      &                                      \\ 
				& Pull    & 1.00              & 0.52              & 1.00              & 0.53              & 1.00              & 0.51              & 1.00              & 0.51              & 1.00              & 0.52              & 1.00              & 0.55              & 1.00             & 0.51             &                                      &                                      \\ 
				\hline
			\end{tabular}
	}
	\caption{\label{tab:siccompare} Peak SIC and ROC AUC achieved when comparing any two samples for each model. CNN, E-flow, and pull represent the three taggers as defined in Figure~\ref{fig:siccurves}. The peak SICs and ROC AUC's which are largest for their combination of samples are shown in bold. The CNN has the strongest performance for all combinations. For all combinations with different final states, the $E$-flow has better performance than pull.}
	\end{table}
\end{center}

\newpage
\subsection{Pearson Correlation Coefficient}

To help understand how jet images are being used to classify the different BSM resonances, the Pearson Correlation Coefficient (PCC) is computed between the true labels and the pixel intensities for each pixel over all images. A PCC image is constructed, where each pixel corresponds to its correlation coefficient. Specifically, let $I_{ij}$  be the distribution of the activations for some pixel $(i,j)$, in the transformed rapidity-azimuthal angle space for every event; let $y$ be the corresponding distribution of true labels. Then the PCC is calculated between these two distributions for that pixel. This is the intensity for that pixel in the new PCC image. 

\begin{figure}[H]
	\centering
	\includegraphics[scale=0.5]{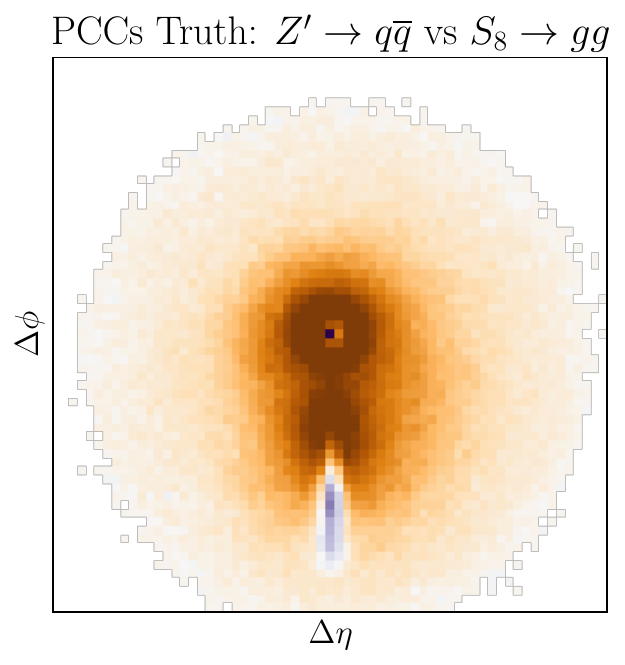}
	\includegraphics[scale=0.5]{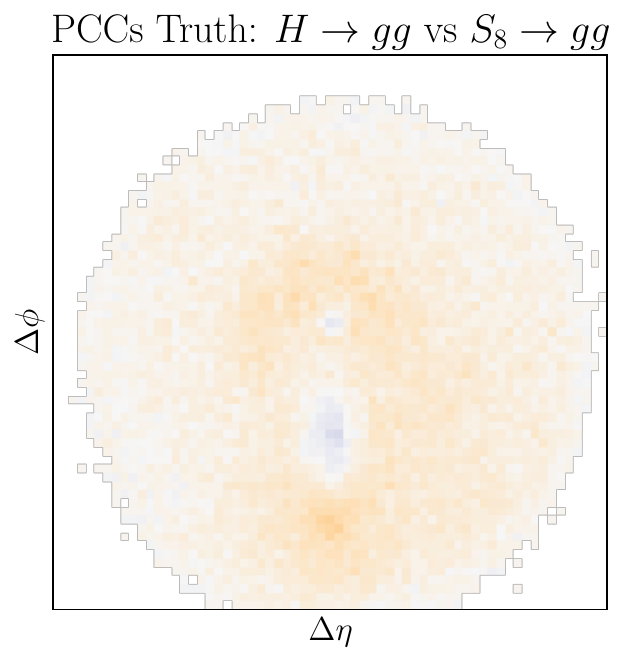}
	\includegraphics[scale=0.5]{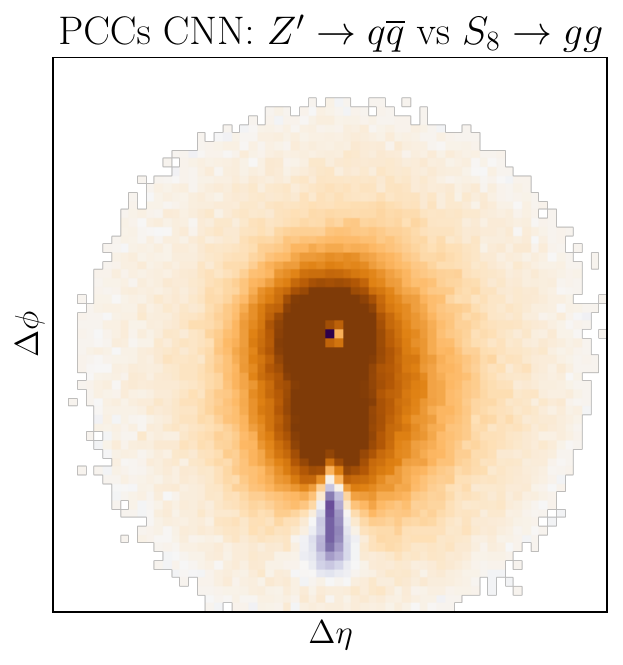}
	\includegraphics[scale=0.5]{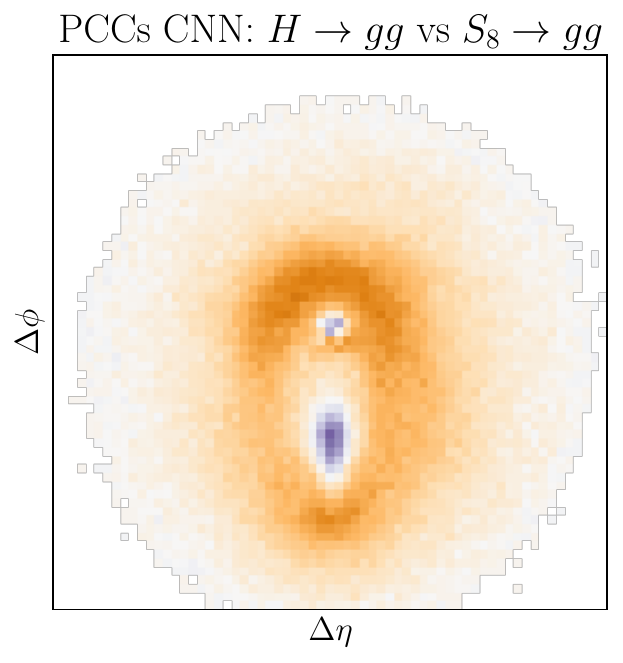}
	\caption{Top: Correlations of the pixel values with true labels. Bottom: Correlation of the pixels with the network labels. Left: PCC image comparing $Z' \rightarrow q\bar{q}$ from $S_8 \rightarrow gg$. Right: PCC image comparing $S_8 \rightarrow gg$ from $H \rightarrow gg$. Orange pixels correlate more the first process; blue pixels correlate more with the second. Note that the processes on the right have the same final state. The correlations of CNN and truth are identical in distribution, with CNN being much stronger.}
	\label{fig:pcccurves}
\end{figure}

PCC images for the two combinations used as examples in Sec.~\ref{sec:ml} are shown in Fig.~\ref{fig:pcccurves}. On the left, there are two orange circles corresponding to the two prongs of the $gg$ dijet, and two blue regions corresponding to the prongs of the $q\bar{q}$ dijet. These regions are a method for visualizing the discriminatory information in each pixel when using a linear model. The PCC images can be used as a computational inspiration for new classifiers. For example, the concentric circles of the E-flows should be effective when they separate the orange regions from the blue regions. On the right, there is a more difficult classification task as the final states are the same. Yet there is still blue and orange regions which could be used as inspirations for new classifiers. For completeness, we provide the PCCs constructed for all combinations of samples are found in the appendix in Fig.~\ref{fig:pcccompare}. Remember that PCC only capture linear correlations between individual pixels and the output, not non-linear correlations or correlations that are functions of multiple pixels. 

\section{\label{sec:dep}Network Dependence on the Setup}
\subsection{Generator Dependence}

The features of the radiation pattern inside jets may not be well-modeled by the simulation.  This is especially true for the case of colorflow information~\cite{Aad:2015lxa,Aaboud:2018ibj}.  While this may be a motivation for an analysis to differentiate various color configurations, it may also have an impact on the quantitative and qualitative conclusions from Sec.~\ref{sec:results}.  The goal of this section is to investigate the impact of simulation variations on the classification performance.  Note that the performance of a network trained on one sample and tested on another may overestimate the impact of a simulation mismodeling - a tagger may identify the correct features even though they may be differently expressed in two different simulations.  This was illustrated for the case of quark/gluon tagging in Ref.~\cite{Komiske:2016rsd}.

Five alternative simulations are prepared to study the modeling dependence.  Three of these samples use the ATLAS A14 tune~\cite{ATL-PHYS-PUB-2014-021} and final state shower variations\footnote{The most important difference between A14 and A14 Var2 is an increase or decrease in the final state shower $\alpha_s$ by about 10\%.} (A14 Var2$\pm$) instead of the default Monash tune~\cite{Skands:2014pea}.  The most important parameter variation in the Var2 tune is the final state shower $\alpha_s$, which is varied by about 10\%.  Two additional samples use alternative color reconnection models.  One of these uses a simple where gluons can be shifted to reduce the total string length (called 'Gluon move')~\cite{Argyropoulos:2014zoa}.  The other sample is more sophisticated and incorporates aspects of QCD (called `QCD inspired')~\cite{Christiansen:2015yqa}.  

\begin{figure}[H]
	\centering
	\includegraphics[scale=0.45]{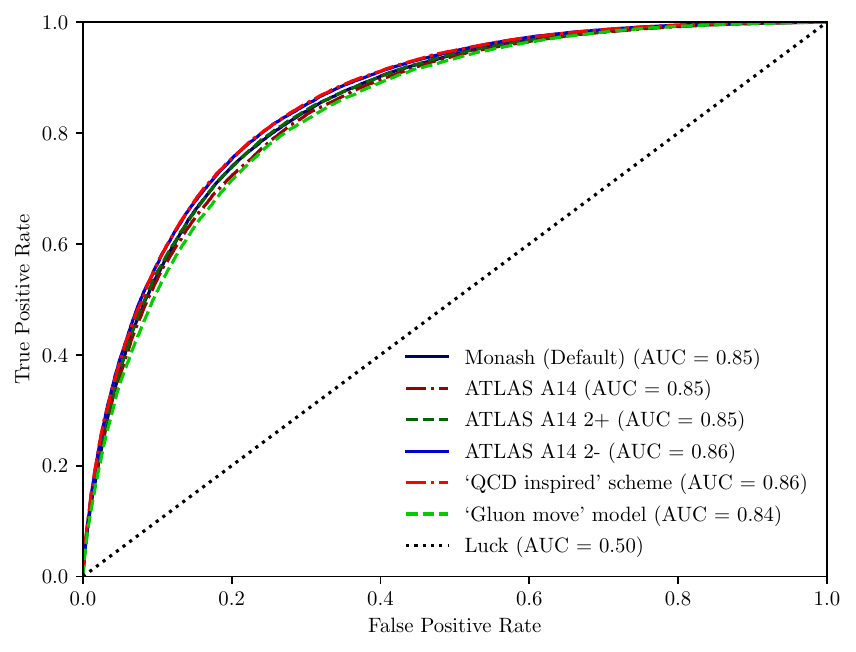}
	\includegraphics[scale=0.45]{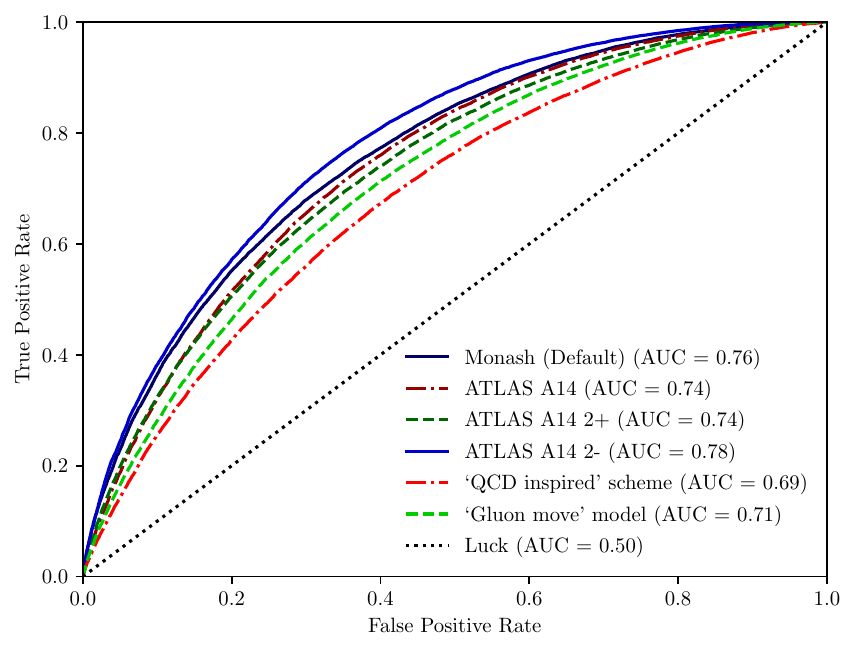}
	\includegraphics[scale=0.45]{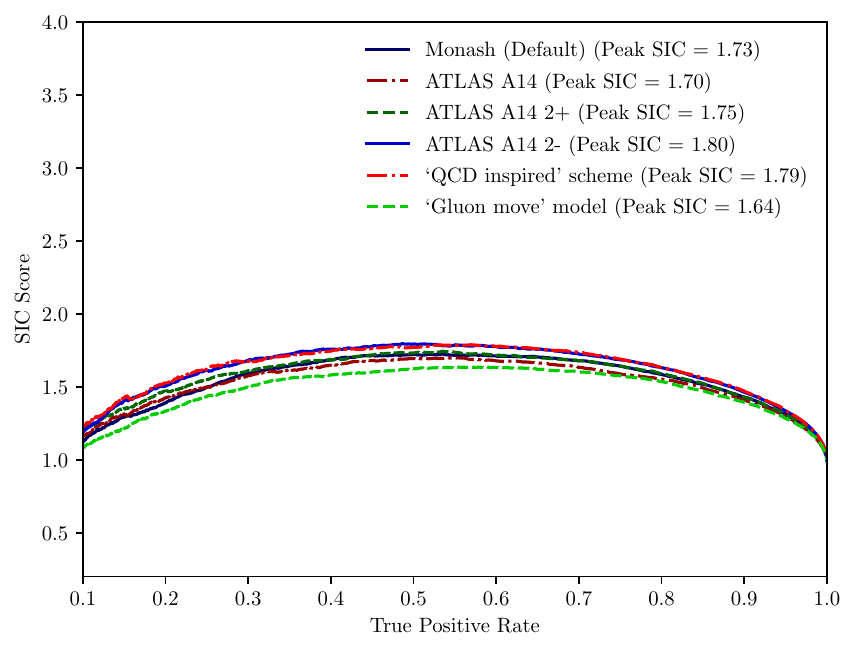}
	\includegraphics[scale=0.45]{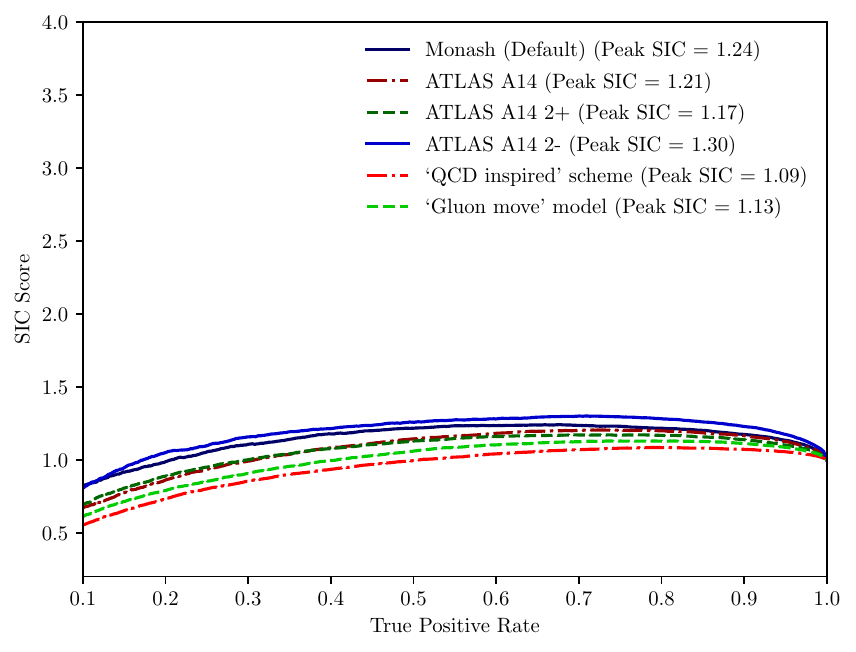}
	\caption{Left: ROC and SIC curves for $C_\mu \rightarrow q\overline{q}$ vs $H \rightarrow gg$. Right: ROC and SIC curves for $C_\mu \rightarrow q\overline{q}$ vs $H \rightarrow q\overline{q}$. All plots contain all different generators that have been checked for generalization of the study.}
	\label{fig:generatorcurves}
\end{figure}

Figure~\ref{fig:generatorcurves} shows the impact of the model variations on the CNN performance.  Overall, the variation is smaller than the spread in performance from Fig.~\ref{fig:siccurves}, which builds confidence in the results from the previous section.  The biggest differences are observed for the variations that modify the final state shower $\alpha_s$.

\subsection{Charged Dependence}

\begin{figure}[H]
	\centering
	\includegraphics[scale=0.49]{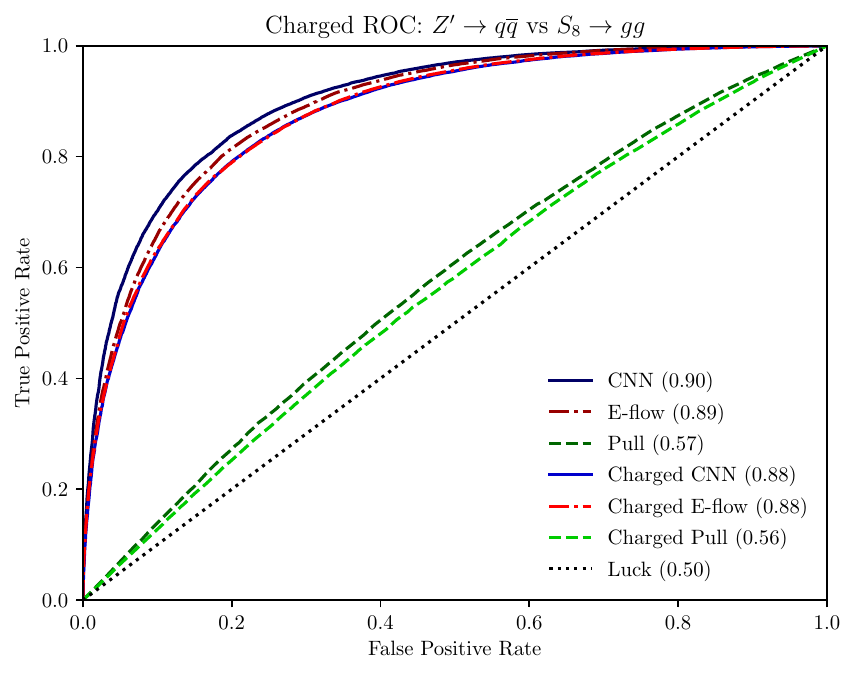}
	\includegraphics[scale=0.49]{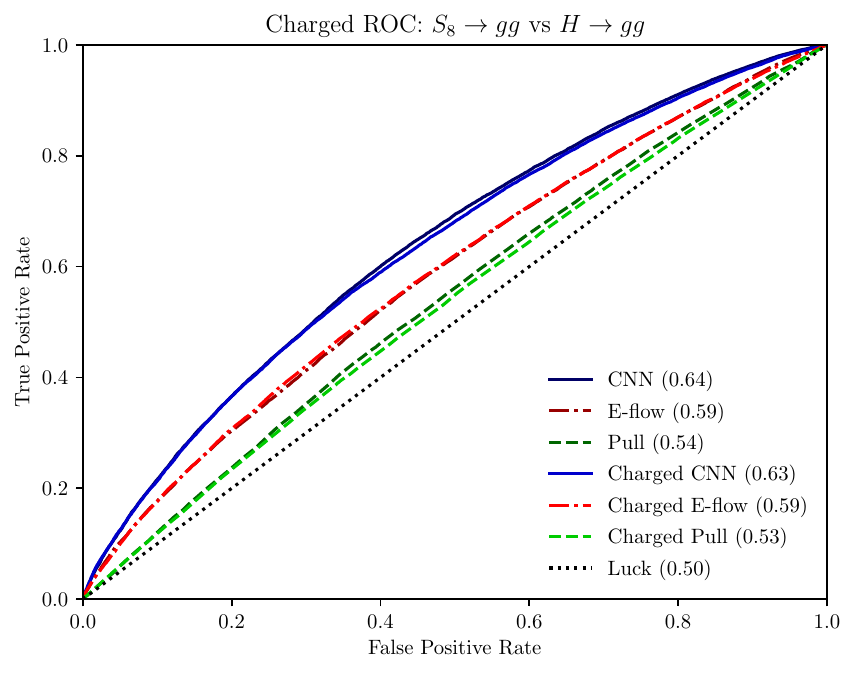}
	\includegraphics[scale=0.49]{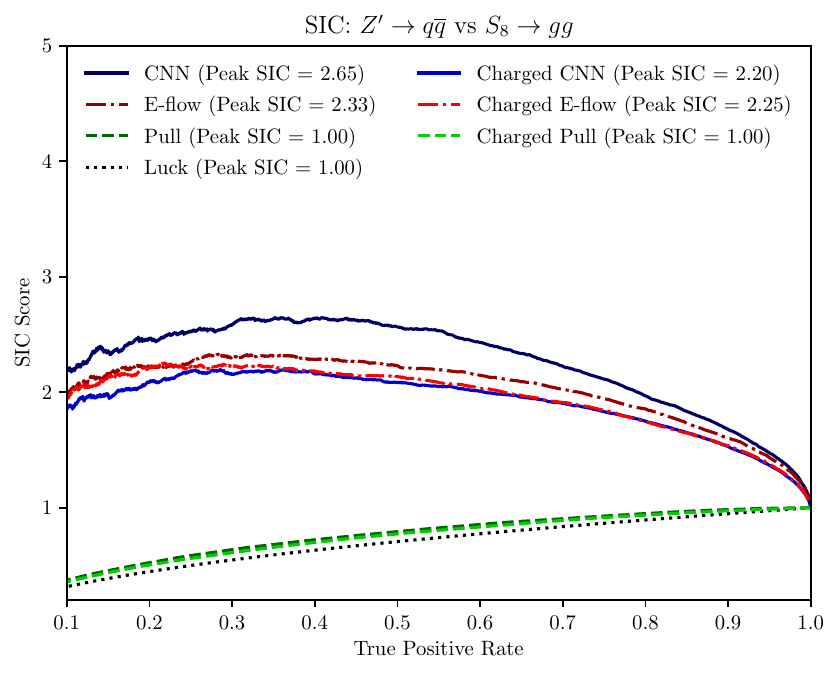}
	\includegraphics[scale=0.49]{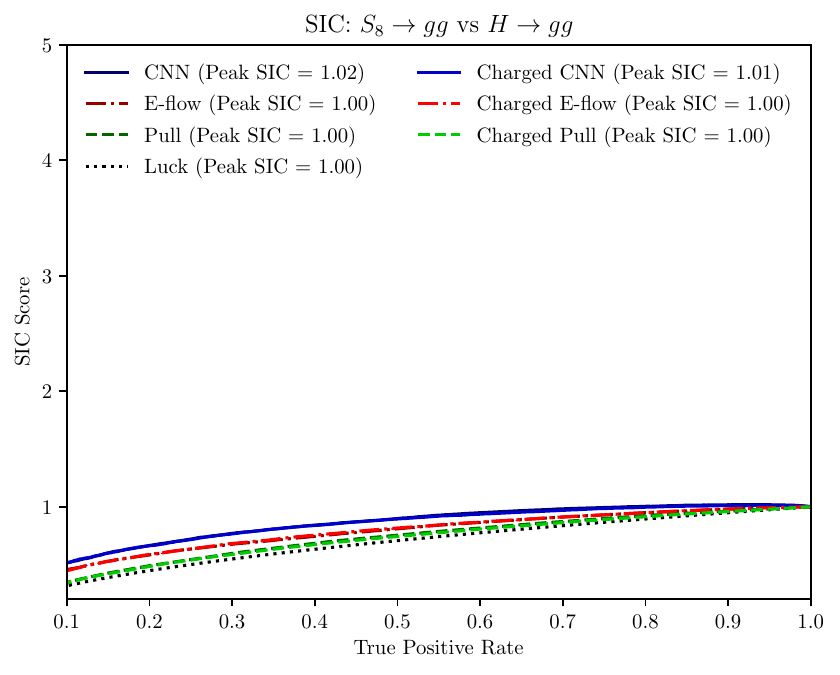}
	\caption{ROC and SIC curves using all-particles (default) or charged-only information for $Z' \rightarrow q\bar{q}$ from $S_8 \rightarrow gg$ and $S_8 \rightarrow gg$ from $H \rightarrow gg$ on the bottom.}
	\label{fig:chargedcurves}
\end{figure}

Thus far, the CNNs have been trained using images built from all final state particles.  However, the energy and position resolution from a real calorimeter may not be suitable for this purpose.  This section explores the impact of only using charged particles, which should be close to realistic information available from charged particle tracks reconstructed from the ATLAS and CMS inner detectors.  Figure~\ref{fig:chargedcurves} compares the all-particles and charged-particles versions of the three taggers considered in Sec.~\ref{sec:results}.  In general, the performance of the charged-only taggers are worse than the all-particles taggers, but the degradation is often smaller than the spread in performance between taggers and the order of tagger performance is preserved.

\subsection{Image Pixelation Dependence}
Various image pixelations of $25\times25$, $50\times50$, $65\times65$, and $100\times100$ were tried.  It was found that increasing image pixelation leads to better results. For example, if performance was measured as the false positive rate (FPR) of the CNN when the true positive rate (TPR) is 0.5, there is a performance increase of $\sim 60 \%$ when increasing image pixelation from $25\times25$ to $50\times50$. However, if the image pixelation is increased from $50\times50$ to $100\times100$, there is only an increase in performance of $\sim 20 \%$. Additionally, larger images require larger networks and take longer to train. The image pixelation of $65\times65$ was chosen as a compromise between speed, network size, and performance. The hyper parameter optimization, as described in \ref{sec:CNNDescription}, was performed after the optimal pixel size was chosen. Future studies that rely on individual jet constituents may be able to benefit from other approaches (see Sec.~\ref{sec:intro}) that do not rely on pixelation.  All ATLAS and CMS detectors are segmented so this benefit will be most pronounced when the pixelation is larger than the detector pixelation as is nearly always the case for tracking detectors.

\subsection{Preprocessing Dependence}

The impact of the preprocessing steps of rotating, normalizing, and logging the images was studied. The FPR of the CNN when the true positive rate TPR is 0.5 was used to measure performance. The rotation step, initially described in~\ref{sec:preML}, did not change performance. Four normalization techniques: $l_1$ normalization, $l_2$ normalization, $Z$-score normalization (also called standardization), and min-max scaling were tried. Each of these on logged images were tried as well. The results are shown in Tab.~\ref{tab:preprocess} below. Based on the FPR values from the table, all jet-images used were rotated, logged, and $l_1$ normalized.
\begin{center}
\begin{table}[!htb]
\resizebox{\textwidth}{!}{
  \begin{tabular}
      {|c|c|c|c|c|c|} \hline 
      & $l_1$ normalization & $l_2$ normalization & Z-score normalization & Min-Max Scaling & No Normalization \\
      \hline 
      No Logging & 0.13 & 0.13 & 0.14 & 0.18 & 0.13\\
      \hline 
      Logging & 0.12 & 0.13 & 0.13 & 0.15 & 0.12\\
       \hline
  \end{tabular}}
  \caption{\label{tab:preprocess} Performance of different pre-processing steps, measured as the FPR of the CNN when the true positive rate TPR is 0.5. The columns show normalization techniques and the rows show whether the images are logged or not.}
\end{table}
\end{center}

 \subsection{Jet Grooming Dependence}
Jets are trimmed by dropping subjets which have a transverse momentum fraction less than $5\%$ of the jet. Grooming the jet in this way has a significant impact on both the CNN and high level taggers. Consider for example comparison of the color singlet $Z^\prime$ with the color octet coloron ($C_\mu$). As mentioned earlier, these two resonances differ only in their color representation. The color flow of their decay products would be different resulting in different radiation patterns. However, grooming the jet would result in some information about these radiation patterns to be lost. On the contrary, we found that the performance of the CNN was better when using trimmed jets than when using untrimmed jets. This suggests that a general approach might not be sufficient and one might need to use a different CNN with possibly a different architecture to pick out these minor differences. Further, it may also happen that there is an optimal transverse momentum fraction cut that would aid in discriminating resonances of different color representation. These considerations are left for future deliberation and work.

\section{\label{sec:concl}Conclusions}

Hypothesized BSM particles that interact with quarks and gluons may be light enough so that they are produced with large transverse momentum at the LHC. When decaying to SM quarks or gluons, the large transverse momentum implies production of a single fat jet. Analyzing the sub structure of this fat jet can help distinguish BSM and SM particles. 

In this paper, we looked at a list of possible BSM particles with a variety of spins, charges and color quantum numbers that decay to pairs of SM particles comprising quarks or gluons. Deep learning was used to improve upon techniques for distinguishing such dijets based on their substructure. Jet images were used to train  a CNN to classify and distinguish between these various types of dijets.
The capabilities of CNNs trained on jet images were compared to other classifiers that used color information, such as Energy Flow Observables and jet pull. Table~\ref{tab:siccompare} demonstrates that the CNN classifier consistently outperforms traditional observables in this task and thus provides a useful method to categorize new particles and expand the jet substructure toolkit.  
The classifier performed well when distinguishing between BSM particles that decay to different final states  (e.g. $Z^{\prime} \to q \bar{q}$ against $H \to g g$), and also did reasonably well when trying to differentiate particles with different spin. On the other hand, its performance was not the strongest when it came to identifying only differences in color quantum numbers. 

Further improvements in these classification tasks might be possible if using a point cloud representation, rather than a CNN. Another avenue that could be explored is the  use of multiclass classification, which would allow us to combine various options into one model without qualitatively affecting performance.

\newpage
\section{\label{sec:app}Appendix}
\begin{center}
\begin{figure}[!htb]
\centering
  \begin{tabular}
      {p{0.5in}p{0.75in}p{0.75in}p{0.75in}p{0.75in}p{0.75in}p{0.75in}p{0.75in}}
      & $Z'\rightarrow q\overline{q}$ & $C_\mu \rightarrow q\overline{q}$ & $H\rightarrow q\overline{q}$ & $q^\star \rightarrow qg$ & $H \rightarrow gg$ & $S_8\rightarrow gg$ & $\Phi^\gamma_6 \rightarrow jj$ \\\\
      $C_\mu \rightarrow q\overline{q}$ &\parbox[c]{10em}{\includegraphics[width=0.75in]{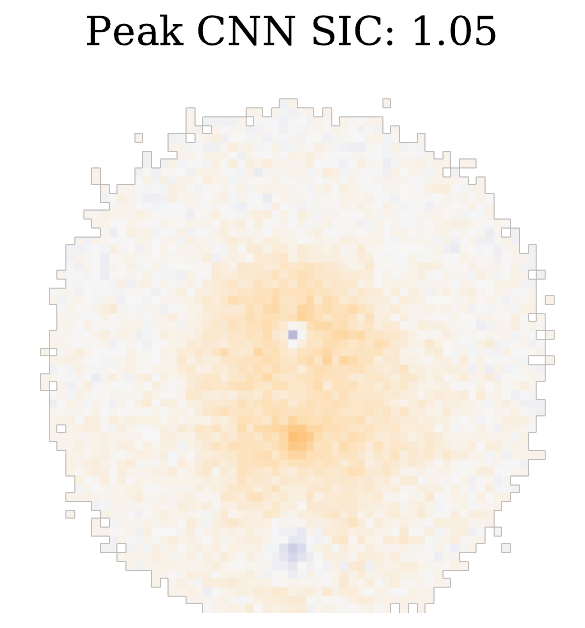}}&&&&&& \\\\
   	  $H\rightarrow q\overline{q}$
      &\parbox[c]{10em}{\includegraphics[width=0.75in]{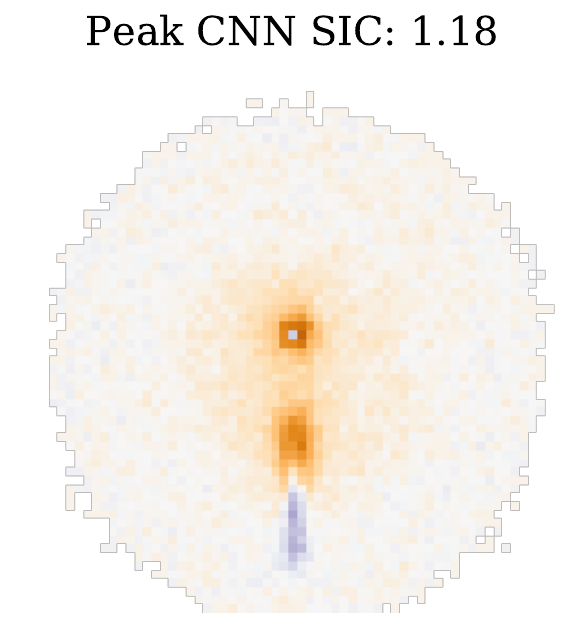}}&\parbox[c]{10em}{\includegraphics[width=0.75in]{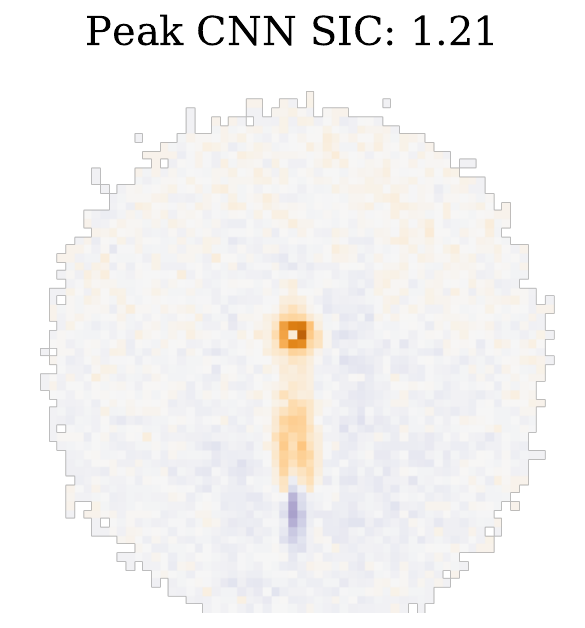}}&&&&& \\\\
      $q^\star \rightarrow qg$
      &\parbox[c]{10em}{\includegraphics[width=0.75in]{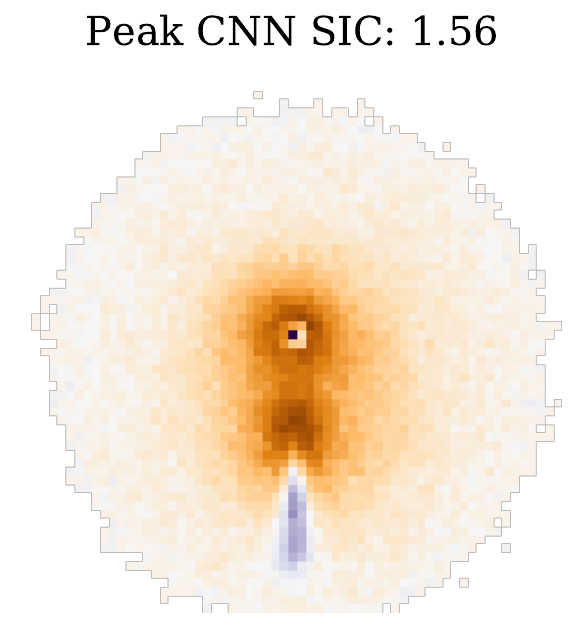}}&\parbox[c]{10em}{\includegraphics[width=0.75in]{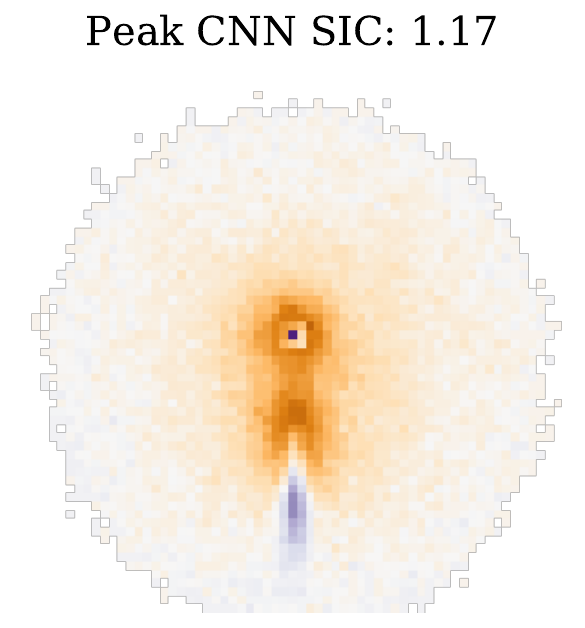}}
      &\parbox[c]{10em}{\includegraphics[width=0.75in]{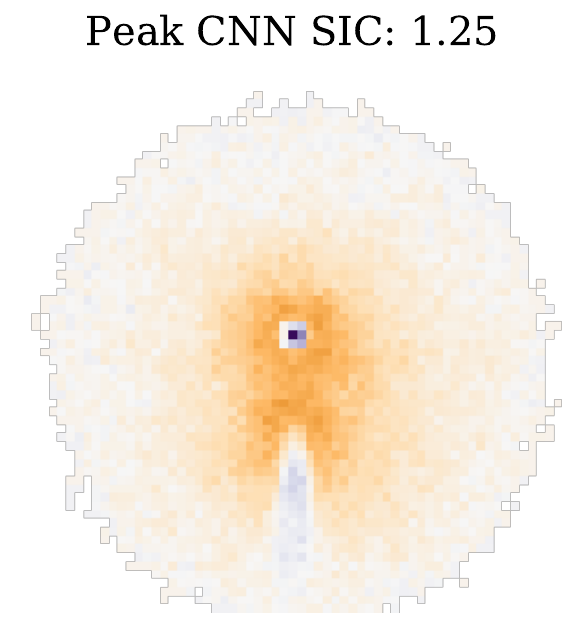}}&&&& \\\\
   	  $H \rightarrow gg$
      &\parbox[c]{10em}{\includegraphics[width=0.75in]{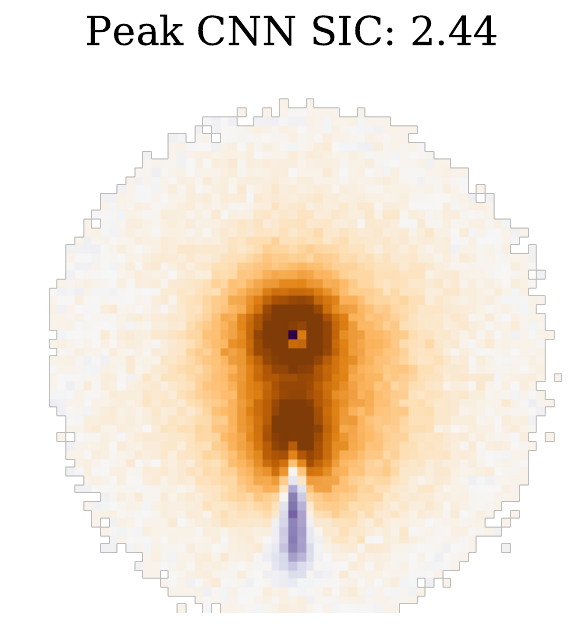}}&\parbox[c]{10em}{\includegraphics[width=0.75in]{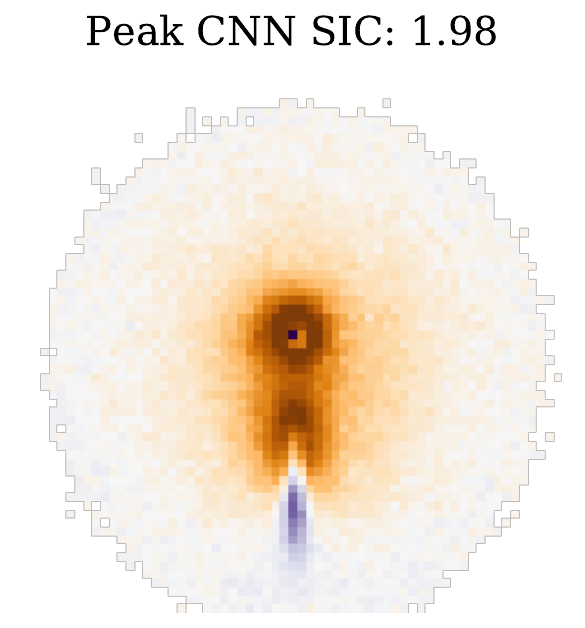}}
      &\parbox[c]{10em}{\includegraphics[width=0.75in]{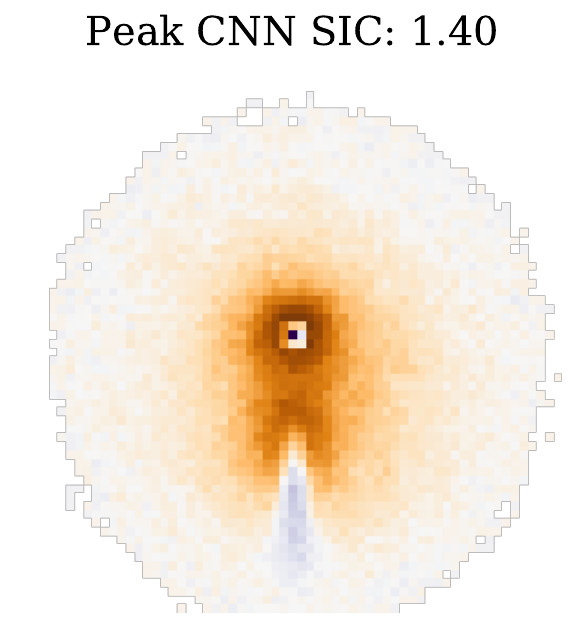}}&\parbox[c]{10em}{\includegraphics[width=0.75in]{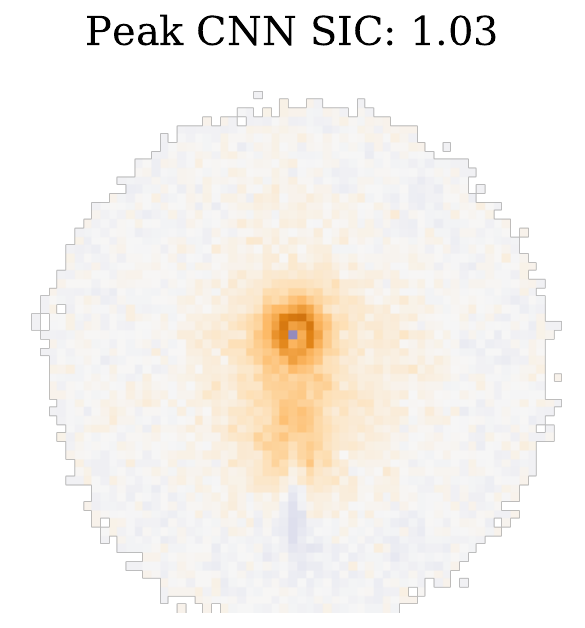}}&&& \\\\
      $S_8\rightarrow gg$ 
      &\parbox[c]{10em}{\includegraphics[width=0.75in]{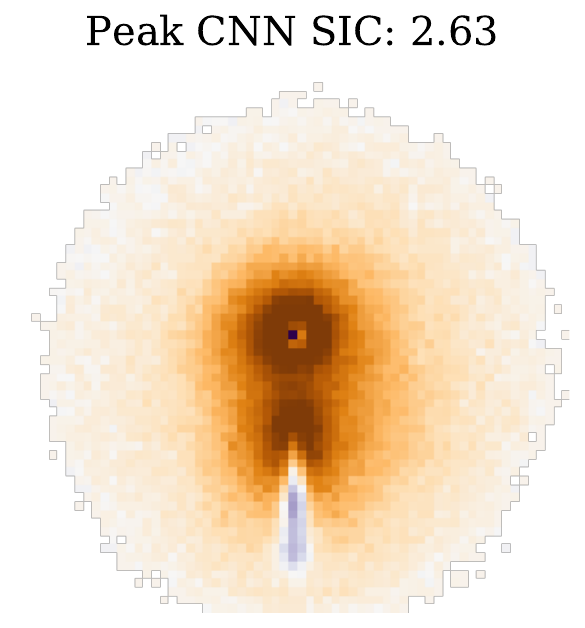}}&\parbox[c]{10em}{\includegraphics[width=0.75in]{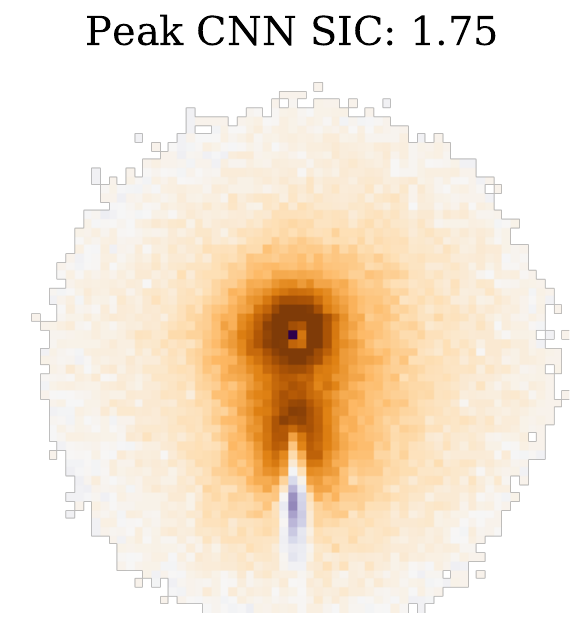}}
      &\parbox[c]{10em}{\includegraphics[width=0.75in]{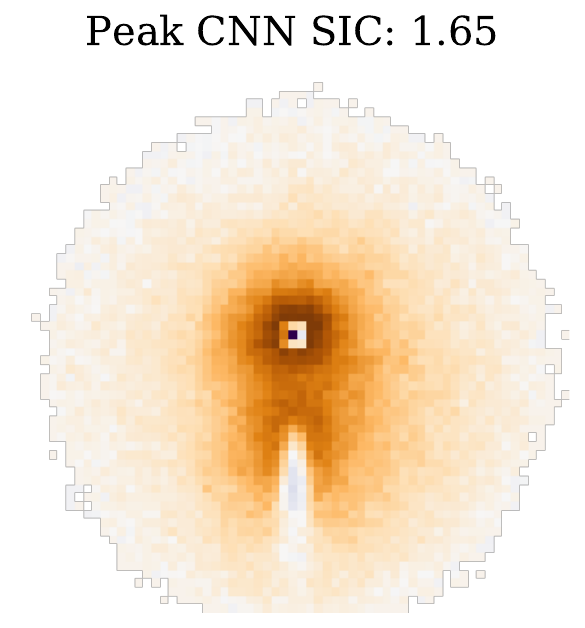}}&\parbox[c]{10em}{\includegraphics[width=0.75in]{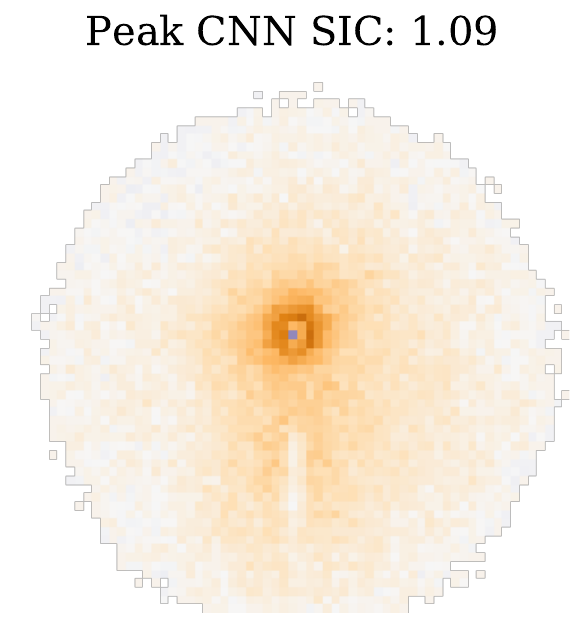}}
      &\parbox[c]{12em}{\includegraphics[width=0.75in]{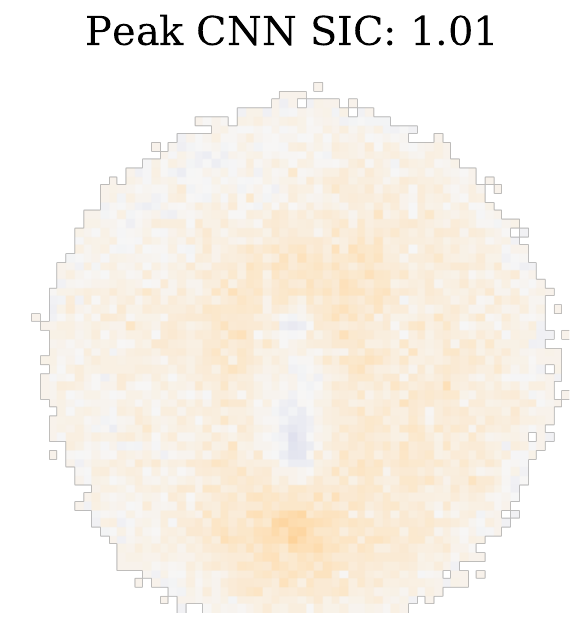}}&& \\
      $\Phi^\gamma_6 \rightarrow jj$
      &\parbox[c]{10em}{\includegraphics[width=0.75in]{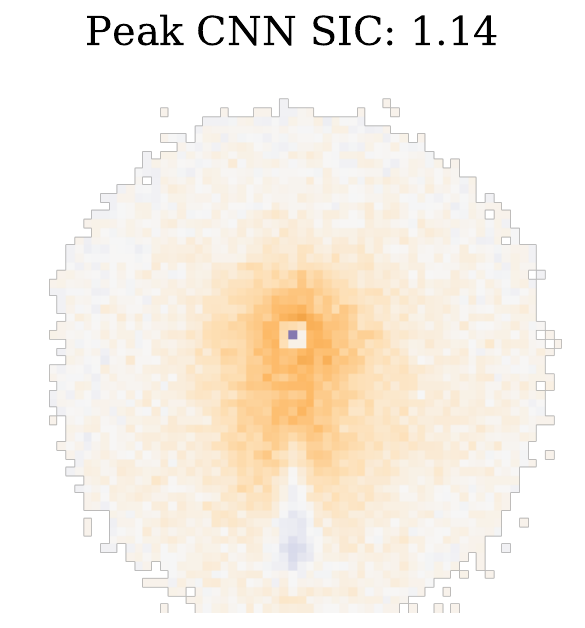}}&\parbox[c]{10em}{\includegraphics[width=0.75in]{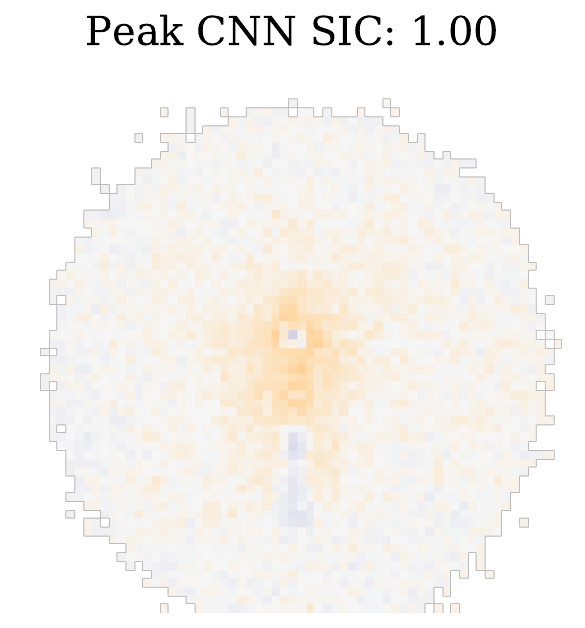}}
      &\parbox[c]{10em}{\includegraphics[width=0.75in]{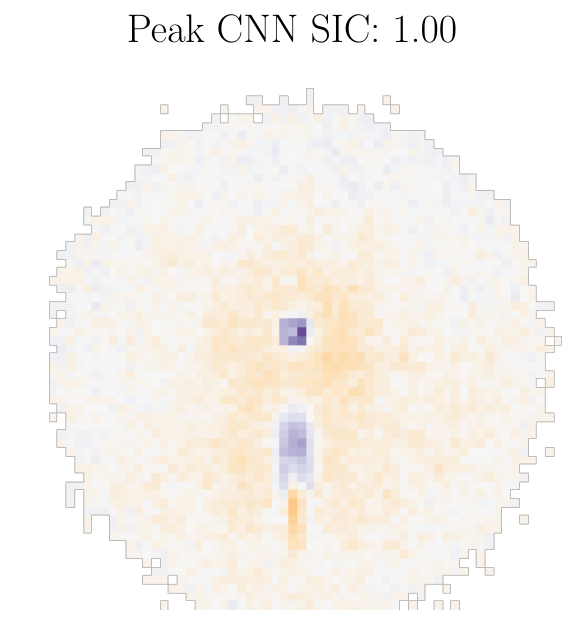}}&\parbox[c]{10em}{\includegraphics[width=0.75in]{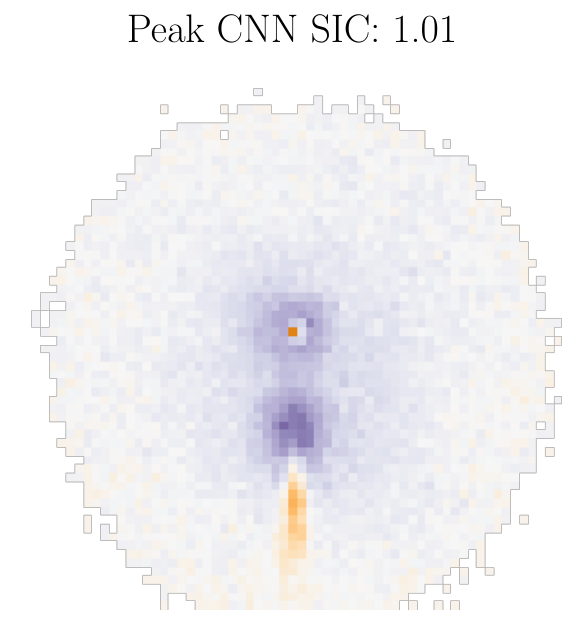}}
      &\parbox[c]{12em}{\includegraphics[width=0.75in]{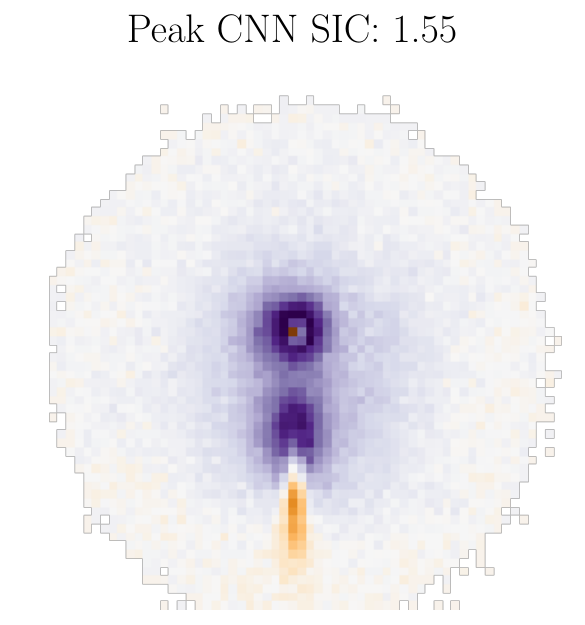}}&\parbox[c]{12em}{\includegraphics[width=0.75in]{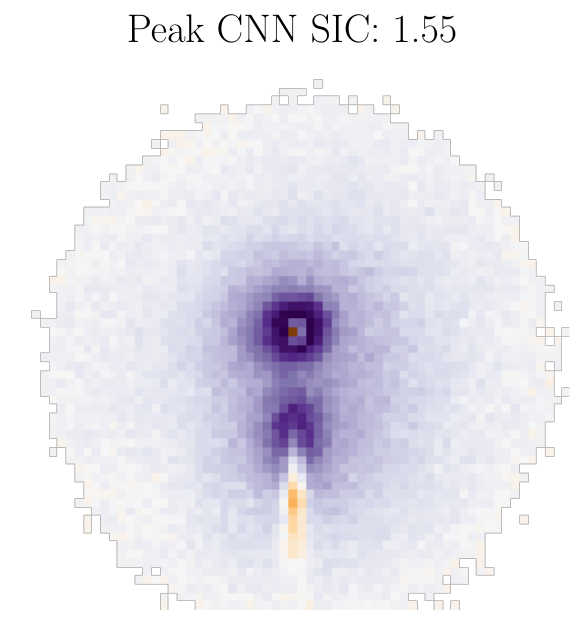}}& \\
	  $X^{\mu\nu} \rightarrow jj$
      &\parbox[c]{10em}{\includegraphics[width=0.75in]{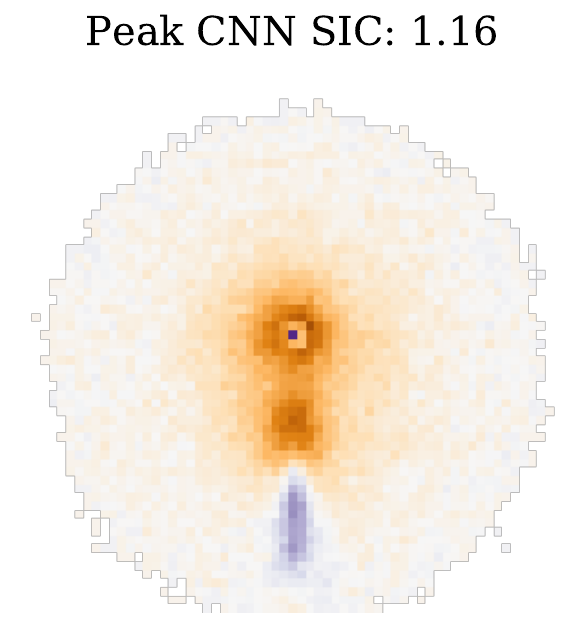}}&\parbox[c]{10em}{\includegraphics[width=0.75in]{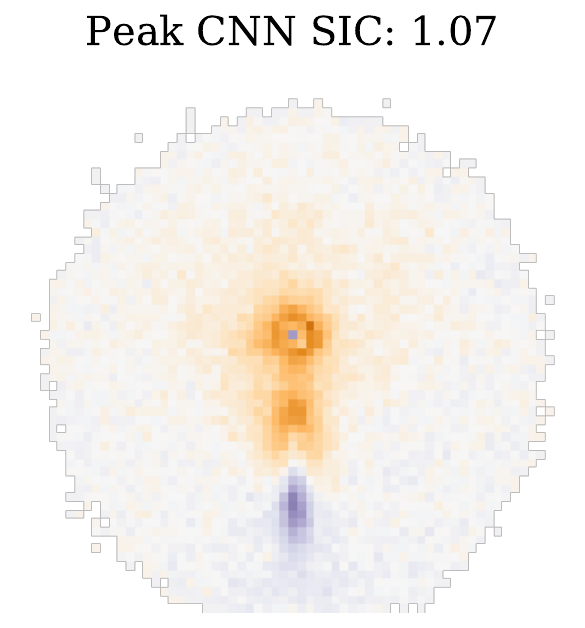}}
      &\parbox[c]{10em}{\includegraphics[width=0.75in]{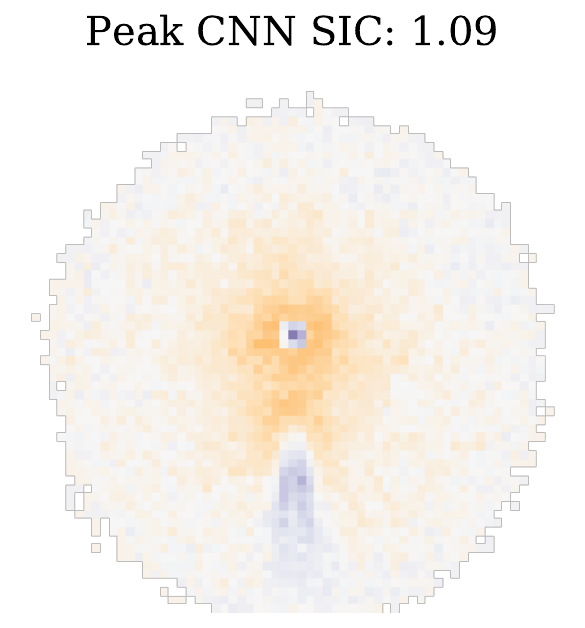}}&\parbox[c]{10em}{\includegraphics[width=0.75in]{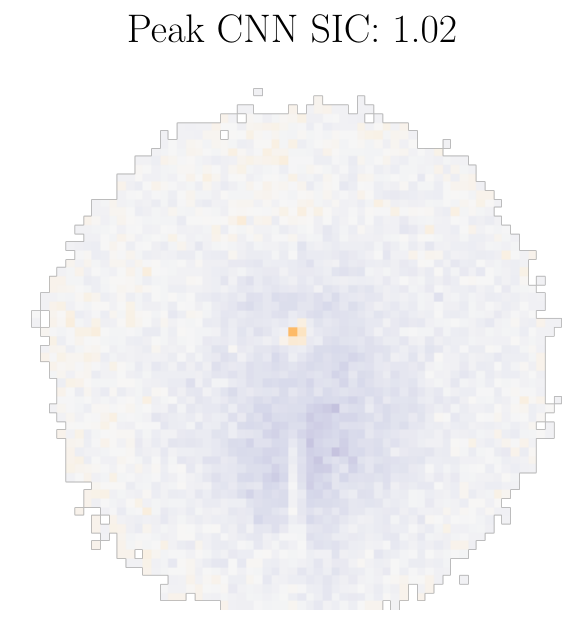}}
      &\parbox[c]{12em}{\includegraphics[width=0.75in]{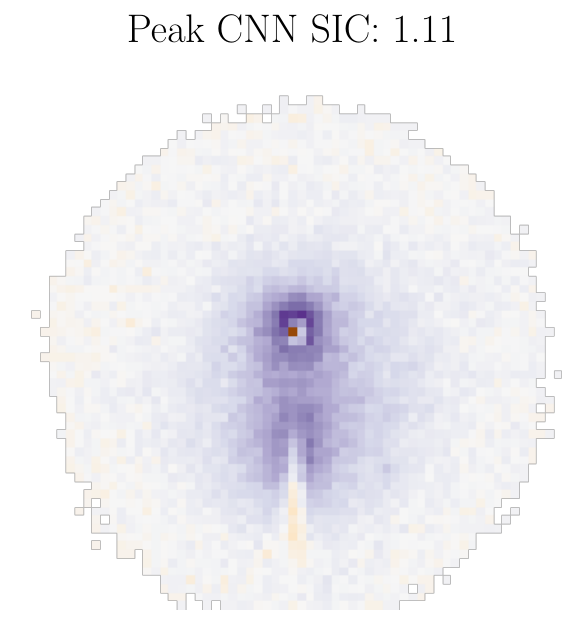}}&\parbox[c]{12em}{\includegraphics[width=0.75in]{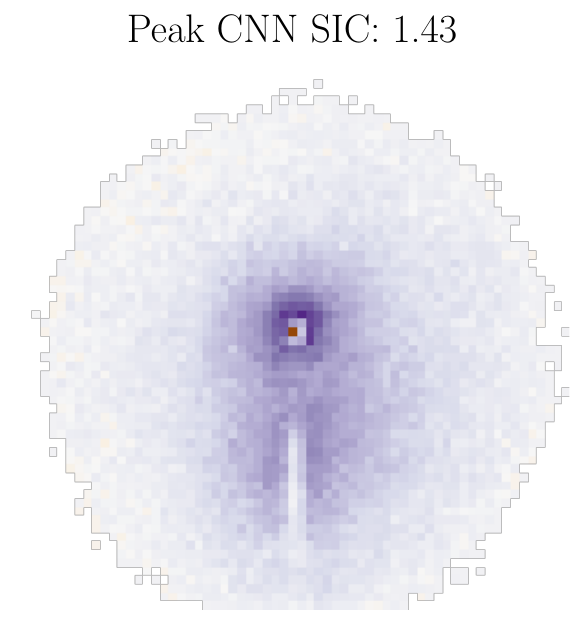}}
      &\parbox[c]{12em}{\includegraphics[width=0.75in]{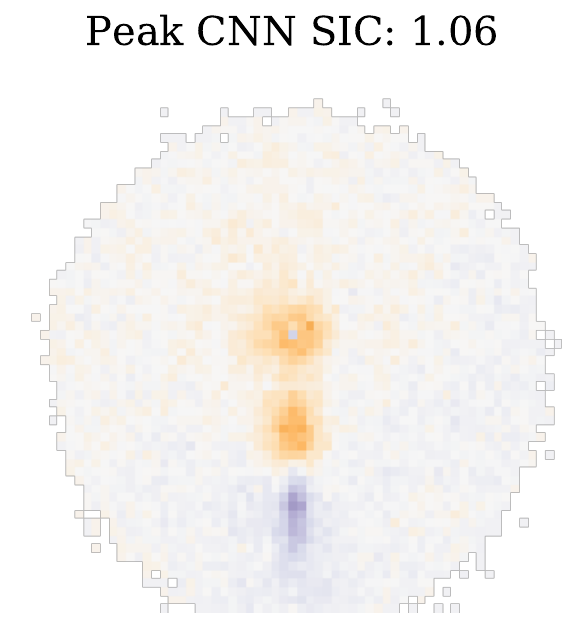}}
      \\
  \end{tabular}
  \caption{\label{fig:pcccompare} All PCC Images for pairwise comparisons.}
\end{figure}
\end{center}

\section*{Acknowledgements}

John Kruper, Jakub Filipek, and Shih-Chieh Hsu were supported by the Department of Energy, Office of High Energy Physics Early Career Research program under Award Number DE-SC0015971. Benjamin Nachman was supported by the U.S. Department of Energy, Office of Science under contract DE-AC02-05CH11231.  BN would like to thank Anton Apostolatos, Leonard Bronner, and Ariel Schwartzman for collaboration at an early stage of this project.
\bibliography{mybib}
\bibliographystyle{jhep}

\end{document}